\documentclass[10pt,journal,compsoc]{IEEEtran}
\usepackage{amsfonts}
\usepackage{amsmath}
\usepackage{cases}
\usepackage{mathrsfs}
\usepackage{bbm}
\usepackage{amsfonts}
\usepackage{color}
\usepackage{multirow}
\usepackage{caption}
\usepackage{flushend}
\usepackage[ruled,vlined,linesnumbered]{algorithm2e}
\usepackage{algorithmic}
\usepackage{subfigure}
\usepackage{extarrows}
\usepackage{graphicx,graphics,txfonts}
\usepackage{array}
\usepackage{tabularx}
\usepackage{stmaryrd}
\usepackage{caption}
\usepackage{multicol}

\begin{document}

\newtheorem{definition}{Definition}
\renewcommand{\algorithmicrequire}{\textbf{Requires}}
\newtheorem{theorem}{Theorem}
\newtheorem{lemma}{Lemma}
\newtheorem{axiom}{Axiom}
\newtheorem{example}{Example}
\newtheorem{corollary}{Corollary}
\newtheorem{property}{Property}

\newcommand{\partitle}[1]{\medskip \noindent \textbf{#1.}}
\newcommand{\subpartitle}[1]{\medskip \emph{#1.}}
\newcommand{\topcaption}{%
\setlength{\abovecaptionskip}{0pt}%
\setlength{\belowcaptionskip}{100pt}%
\caption}

\title{Secure and Efficient Skyline Queries on Encrypted Data}

\author{Jinfei Liu,~\IEEEmembership{member,~IEEE,}
        Juncheng Yang,~\IEEEmembership{member,~IEEE,}
        Li Xiong,~\IEEEmembership{member,~IEEE,}
        and Jian Pei,~\IEEEmembership{Fellow,~IEEE}
\IEEEcompsocitemizethanks{
\IEEEcompsocthanksitem Jinfei Liu, Juncheng Yang, and Li
Xiong are with the Department of Mathematics and
Computer Science, Emory University.\protect\\
E-mail: \{jinfei.liu, juncheng.yang, and lxiong\}@emory.edu}

\IEEEcompsocitemizethanks{ \IEEEcompsocthanksitem Jian Pei is with
School of Computing Science, Simon Fraser University.\protect\\
E-mail: jpei@cs.sfu.ca}

\thanks{Manuscript received XXXXXX; revised XXXXXX.}
}


\IEEEtitleabstractindextext{
\begin{abstract}
Outsourcing data and computation to cloud server provides a cost-effective way to support large scale data storage and query processing. However, due to security and privacy concerns, sensitive data (e.g., medical records) need to be protected from the cloud server and other unauthorized users. One approach is to outsource encrypted data to the cloud server and have the cloud server perform query processing on the encrypted data only. It remains a challenging task to support various queries over encrypted data in a secure and efficient way such that the cloud server does not gain any knowledge about the data, query, and query result. In this paper, we study the problem of secure skyline queries over encrypted data. The skyline query is particularly important for multi-criteria decision making but also presents significant challenges due to its complex computations. We propose a fully secure skyline query protocol on data encrypted using semantically-secure encryption. As a key subroutine, we present a new secure dominance protocol, which can be also used as a building block for other queries. Furthermore, we demonstrate two optimizations, data partitioning and lazy merging, to further reduce the computation load. Finally, we provide both serial and parallelized implementations and empirically study the protocols in terms of efficiency and scalability under different parameter settings, verifying the feasibility of our proposed solutions.
\end{abstract}

\begin{IEEEkeywords}
Skyline, Secure, Efficient, Parallel, Semi-honest.
\end{IEEEkeywords}
}

\maketitle
\section{Introduction}\label{sec:Introduction}

As an emerging computing paradigm, cloud computing attracts increasing attention from both research and industry communities. Outsourcing data and computation to cloud server provides a cost-effective way to support large scale data storage and query processing. However, due to security and privacy concerns, sensitive data need to be protected from the cloud server as well as other unauthorized users.

\begin{figure}[htb]
 \centering
 \includegraphics[width=0.45\textwidth]{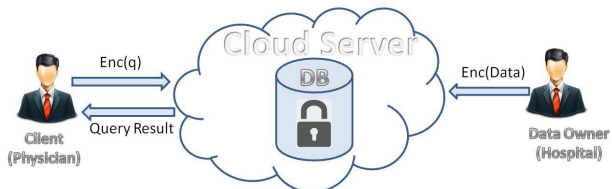}
 \vspace{-1em}
 \caption{Secure similarity queries.}
 \label{fig:secure}
\end{figure}
 \vspace{-1em}

A common approach to protect the confidentiality of outsourced data is to encrypt the data (e.g., \cite{DBLP:conf/stoc/Gentry09,DBLP:conf/eurocrypt/Paillier99}). To protect the confidentiality of the query from cloud server, authorized clients also send encrypted queries to the cloud server. Figure \ref{fig:secure} illustrates our problem scenario of secure query processing over encrypted data in the cloud. The data owner outsources encrypted data to the cloud server.  The cloud server processes encrypted queries from the client on the encrypted data and returns the query result to the client. During the query processing, the cloud server should not gain any knowledge about the data, data patterns, query, and query result.

Fully homomorphic encryption schemes \cite{DBLP:conf/stoc/Gentry09} ensure strong security while enabling arbitrary computations on the encrypted data. However, the computation cost is prohibitive in practice. Trusted hardware such as Intel's Software Guard Extensions (SGX) brings a promising alternative, but still has limitations in its security guarantees \cite{costanintel}. Many techniques (e.g., \cite{DBLP:conf/sigmod/HacigumusILM02,DBLP:conf/sp/SongWP00}) have been proposed to support specific queries or computations on encrypted data with varying degrees of security guarantee and efficiency (e.g., by weaker encryptions). Focusing on similarity search, secure $k$-nearest neighbor ($k$NN) queries, which return $k$ most similar (closest) records given a query record, have been extensively studied \cite{DBLP:conf/icde/ElmehdwiSJ14,DBLP:conf/icde/HuXRC11,DBLP:conf/sigmod/WongCKM09,DBLP:conf/icde/0002LX13}.

In this paper, we focus on the problem of secure skyline queries on encrypted data, another type of similarity search important for multi-criteria decision making. The {\em skyline} or {\em Pareto} of a multi-dimensional dataset given a query point consists of the data points that are not {\em dominated} by other points.  A data point dominates another if it is closer to the query point in at least one dimension and at least as close to the query point in every other dimension.  The skyline query is particularly useful for selecting similar (or best) records when a single aggregated distance metric with all dimensions is hard to define.  The assumption of $k$NN queries is that the relative weights of the attributes are known in advance, so that a single similarity metric can be computed between a pair of records aggregating the similarity between all attribute pairs. However, this assumption does not always hold in practical applications. In many scenarios, it is desirable to retrieve similar records considering all possible relative weights of the attributes (e.g., considering only one attribute, or an arbitrary combination of attributes), which is essentially the skyline or the ``pareto-similar" records.

\begin{figure}[htb]
 \centering
 \includegraphics[width=0.28\textwidth]{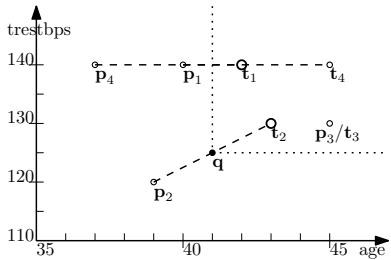}
 \caption{Dynamic skyline query.}
 \label{fig:initEG}
\end{figure}

\partitle{Motivating Example}
Consider a hospital who wishes to outsource its electronic health records to the cloud and the data is encrypted to ensure data confidentiality. Let $P$ denote a sample heart disease dataset with attributes ID, age, trestbps (resting blood pressure). We sampled four patient records $\textbf{p}_1,...,\textbf{p}_4$ from the heart disease dataset of UCI machine learning repository \cite{DBLP:ucidata} as shown in Table \ref{tab:heartD}(a) and Figure \ref{fig:initEG}. Consider a physician who is treating a heart disease patient $\textbf{q}=(41,125)$ and wishes to retrieve similar patients in order to enhance and personalize the treatment for patient $\textbf{q}$. While it is unclear how to define the attribute weights for $k$NN queries ($\textbf{p}_1$ is the nearest if only age is considered while $\textbf{p}_2,\textbf{p}_3$ are the nearest if only trestbps is considered), skyline provides all pareto-similar records that are not dominated by any other records. Skyline includes all possible 1NN results by considering all possible relative attribute weights, and hence can serve as a filter for users. Given the query $\textbf{q}$, we can map the data points to a new space with $\textbf{q}$ as the origin and the distance to $\textbf{q}$ as the mapping function. The mapped records $\textbf{t}_i[j]=|\textbf{p}_i[j]-\textbf{q}[j]|+\textbf{q}[j]$ on each dimension $j$ are shown in Table \ref{tab:heartD}(b) and also in Figure \ref{fig:initEG}. It is easy to see that $\textbf{t}_1$ and $\textbf{t}_2$ are skyline in the mapped space, which means $\textbf{p}_1$ and $\textbf{p}_2$ are skyline with respect to query $\textbf{q}$.

Our goal is for the cloud server to compute the skyline query given $\textbf{q}$ on the encrypted data without revealing the data, the query $\textbf{q}$, the final result set $\{\textbf{p}_1, \textbf{p}_2\}$, as well as any intermediate result (e.g., $\textbf{t}_2$ dominates $\textbf{t}_4$) to the cloud. We note that skyline computation (with query point at the origin) is a special case of skyline queries.  

\begin{table}[h]
\caption{Sample of heart disease dataset.}\footnotesize\label{tab:heartD}
\vspace{-1em}
\centering
\subtable[Original data.]{
\begin{tabular}{|c|c|c|}
\hline
ID & age & trestbps\\
\hline
$\textbf{p}_1$ & 40 & 140\\
\hline
$\textbf{p}_2$ & 39 & 120\\
\hline
$\textbf{p}_3$ & 45 & 130\\
\hline
$\textbf{p}_4$ & 37 & 140\\
\hline
\end{tabular}
}
\qquad
\subtable[Mapped Data.]{
\begin{tabular}{|c|c|c|}
\hline
ID & age & trestbps\\
\hline
$\textbf{t}_1$ & 42 & 140\\
\hline
$\textbf{t}_2$ & 43 & 130\\
\hline
$\textbf{t}_3$ & 45 & 130\\
\hline
$\textbf{t}_4$ & 45 & 140\\
\hline
\end{tabular}
       \label{tab:secondtable}
}
\end{table}

\vspace{-2em}
\vspace{-2em}

\partitle{Challenges}
Designing a fully secure protocol for skyline queries over encrypted data presents significant challenges due to the complex comparisons and computations. Let $P$ denotes a set of $n$ tuples $\textbf{p}_1,...,\textbf{p}_n$ with $m$ attributes and $\textbf{q}$ denotes input query tuple.  In $k$NN queries, we only need to compute the distances between each tuple $\textbf{p}_i$ and the query tuple $\textbf{q}$ and then choose the $k$ tuples corresponding to the $k$ smallest distances.  In skyline queries, for each tuple $\textbf{p}_i$, we need to compare it with all other tuples to check dominance. For each comparison between two tuples $\textbf{p}_a$ and $\textbf{p}_b$, we need to compare all their $m$ attributes and for comparison of each attribute $\textbf{p}[j]$, there are three different outputs, i.e., $\textbf{p}_a[j]<(=,>)$  $\textbf{p}_b[j]$. Therefore, there are $3^m$ different outputs for each comparison between two tuples, based on which we need to determine if one tuple dominates the other. How to determine the $2^m-1$ cases that satisfy $p_a$ dominates $p_b$ efficiently while protecting intermediate results (e.g., whether two attribute values are the same) is particularly challenging.

Such complex comparisons and computations require more complex protocol design in order to carry out the computations on the encrypted data given an encryption scheme with semantic security (instead of weaker order-preserving or other property-preserving encryptions).  In addition, the extensive intermediate result means more indirect information about the data can be potentially revealed (e.g., which tuple dominates which other, whether there are duplicate tuples or equivalent attribute values) even if the exact data is protected.  This makes it challenging to design a fully secure and efficient skyline query protocol in which the cloud should not gain any knowledge about the data including indirect data patterns.
\vspace{-0.5em}

\partitle{Contributions}
We summarize our contributions as follows.
\vspace{-0.5em}

\begin{itemize}
\item We study the secure skyline problem on encrypted data with semantic security for the first time.  We assume the data is encrypted using the Paillier cryptosystem which provides semantic security and is partially homomorphic.

\item We propose a fully secure dominance protocol, which can be used as a building block for skyline queries as well as other queries, e.g., reverse skyline queries \cite{DBLP:conf/vldb/DellisS07} and $k$-skyband queries \cite{DBLP:journals/tods/PapadiasTFS05}.

\item We present two secure skyline query protocols. The first one, served as a basic and efficient solution, leaks some indirect data patterns to the cloud server. The second one is fully secure and ensures that the cloud gains no knowledge about the data including indirect patterns. The proposed protocols exploit the partial (additive) homomorphism as well as novel permutation and perturbation techniques to ensure the correct result is computed while guaranteeing privacy. We provide security and complexity analysis of the proposed protocols.

\item Compared with our conference version \cite{DBLP:conf/icde/LiuY0P17}, we present two new optimizations, data partitioning and lazy merging, to further reduce the computation load. For the data partitioning, we theoretically analyze the optimal number of partitions given the number of points, the expected number of output skyline points, the number of decomposed bits, and the number of dimensions. In addition, we propose a lazy merging scheme that aims to reduce computation overhead due to the smaller partition sizes at the later stage of the partitioning scheme.

\item We also provide a complete implementation including both serial and parallelized versions which can be deployed in practical settings. We empirically study the efficiency and scalability of the implementations under different parameter settings, verifying the feasibility of our proposed solutions.
\end{itemize}
\vspace{-0.5em}

\partitle{Organization}
The rest of the paper is organized as follows. Section \ref{sec:relatedwork} presents the related work. Section \ref{sec:pre} introduces background definitions as well as our problem setting. The security subprotocols for general functions that will be used in our secure skyline protocol are introduced in Section \ref{sec:primitives}. The key subroutine of secure skyline protocols, secure dominance protocol, is shown in Section \ref{sec:secDom}. The complete secure skyline protocols are presented in Section \ref{sec:SSkylinePro}. We illustrate two optimizations to further reduce the computation load in Section \ref{sec:Perfor}. We report the experimental results and findings in Section \ref{sec:experiments}. Section \ref{sec:conclusion} concludes the paper.
\section{Related Work}\label{sec:relatedwork}

\partitle{Skyline}
The skyline computation problem was first studied in computational geometry field \cite{DBLP:journals/cacm/Bentley80, DBLP:journals/jacm/KungLP75} where they focused on worst-case time complexity. \cite{DBLP:conf/compgeom/KirkpatrickS85, DBLP:journals/ipl/LiuXX14} proposed output-sensitive algorithms achieving $O(nlogk)$ in worst-case where $k$ is the number of skyline points which is far less than $n$ in general.

Since the introduction of the skyline operator by B\"{o}rzs\"{o}nyi et al. \cite{DBLP:conf/icde/BorzsonyiKS01}, skyline has been extensively studied in the database field. Kossmann et al. \cite{DBLP:conf/vldb/KossmannRR02} studied the progressive algorithm for skyline queries. Different variants of the skyline problem have been studied (e.g., subspace skyline \cite{DBLP:conf/sigmod/ChanJTTZ06}, uncertain skyline \cite{DBLP:conf/vldb/PeiJLY07} \cite{DBLP:conf/cikm/LiuZXLL15}, group-based skyline \cite{DBLP:journals/pvldb/LiuXPLZ15, DBLP:conf/cikm/Li0HRD12, DBLP:conf/cikm/YuQL0CZ17}, skyline diagram \cite{DBLP:conf/icde/LiuY0PL18}).

\partitle{Secure query processing on encrypted data}
Fully homomorphic encryption schemes \cite{DBLP:conf/stoc/Gentry09} enable arbitrary computations on encrypted data. Even though it is shown that \cite{DBLP:conf/stoc/Gentry09} we can build such encryption schemes with polynomial time, they remain far from practical even with the state of the art implementations \cite{DBLP:conf/eurocrypt/HaleviS15}.

Many techniques (e.g., \cite{DBLP:conf/sigmod/HacigumusILM02,DBLP:conf/sp/SongWP00}) have been proposed to support specific queries or computations on encrypted data with varying degrees of security guarantee and efficiency (e.g., by weaker encryptions).
We are not aware of any formal work on secure skyline queries over encrypted data with semantic security.  Bothe et al. \cite{DBLP:conf/cikm/BotheCKV14} and their demo version \cite{DBLP:journals/pvldb/BotheKV13} illustrated an approach about skyline queries on so-called ``encrypted'' data without any formal security guarantee.  Another work \cite{DBLP:conf/infocom/ChenLZZL16} studied the verification of skyline query result returned by an untrusted cloud server.

The closely related work is secure $k$NN queries \cite{DBLP:conf/icde/ElmehdwiSJ14,DBLP:conf/edbt/HashemKZ10,DBLP:conf/icde/HuXRC11,DBLP:journals/pvldb/PapadopoulosBP10,DBLP:conf/icdcs/QiA08,DBLP:conf/sigmod/WongCKM09,DBLP:conf/icde/0002LX13,DBLP:conf/icde/YiPBV14} which we discuss in more detail here.
Wong et al. \cite{DBLP:conf/sigmod/WongCKM09} proposed a new encryption scheme called asymmetric scalar-product-preserving encryption. In their work, data and query are encrypted using slightly different encryption schemes and all clients know the private key. Hu et al. \cite{DBLP:conf/icde/HuXRC11} proposed a method based on provably secure privacy homomorphism encryption scheme. However, both schemes are vulnerable to the chosen-plaintext attacks as illustrated by Yao et al. \cite{DBLP:conf/icde/0002LX13}. Yao et al. \cite{DBLP:conf/icde/0002LX13} proposed a new method based on secure Voronoi diagram. Instead of asking the cloud server to retrieve the exact $k$NN result, their method retrieve a relevant encrypted partition such that it is guaranteed to contain the kNN of the query point. Hashem et al. \cite{DBLP:conf/edbt/HashemKZ10} identified the challenges in preserving user privacy for group nearest neighbor queries and provided a comprehensive solution to this problem. Yi et al. \cite{DBLP:conf/icde/YiPBV14} proposed solutions for secure $k$NN queries based on oblivious transfer paradigm. Recently, Elmehdwi et al. \cite{DBLP:conf/icde/ElmehdwiSJ14} proposed a secure $k$NN query protocol on data encrypted using Paillier cryptosystem that ensures data privacy and query privacy, as well as low (or no) computation overhead on client and data owner using two non-colluding cloud servers. Our work follows this setting and addresses skyline queries.

Other works studied $k$NN queries in the secure multi-party computation (SMC) setting \cite{DBLP:conf/icdcs/QiA08} (data is distributed between two parties who want to cooperatively compute the answers without revealing to each other their private data), or private information retrieval (PIR) setting \cite{DBLP:journals/pvldb/PapadopoulosBP10} (query is private while data is public), which are different from our settings.

\partitle{Secure Multi-party Computation (SMC)} SMC was first proposed by Yao \cite{DBLP:conf/focs/Yao82b} for two-party setting and then extended by Goldreich et al. \cite{DBLP:conf/stoc/GoldreichMW87} to multi-party setting. SMC refers to the problem where a set of parties with private inputs wish to compute some joint function of their inputs. There are techniques such as garbled circuits \cite{DBLP:conf/uss/HuangEKM11} and secret sharing \cite{beimel2011secret} that can be used for SMC. In this paper, all protocols assume a two-party setting, but different from the traditional SMC setting. Namely, we have party $\mathcal{C}_1$ with encrypted input and party $\mathcal{C}_2$ with the private key $sk$. The goal is for $\mathcal{C}_1$ to obtain an encrypted result of a function on the input without disclosing the original input to either $\mathcal{C}_1$ or $\mathcal{C}_2$.

\section{Preliminaries and Problem Definitions}\label{sec:pre}

In this section, we first illustrate some background knowledge on skyline computation and dynamic skyline query, and then describe the security model we use in this paper. For references, a summary of notations is given in Table \ref{tab:notations}.

\begin{table}[h]\centering
\caption{The summary of notations.}\vspace{-1em}\footnotesize\label{tab:notations}
{%
\begin{tabular}{|c|c|}
\hline
Notation & Definition\\
\hline
$P$   & dataset of $n$ points/tuples/records\\
\hline
$\textbf{p}_i[j]$ & the $j^{th}$ attribute of $\textbf{p}_i$\\
\hline
$\textbf{q}$   & query tuple of client\\
\hline
$n$ & number of points in $P$\\
\hline
$m$ & number of dimensions\\
\hline
$k$ & number of skyline\\
\hline
$l$ & number of bits\\
\hline
$K$ & key size\\
\hline
$pk/sk$ & public/private key\\
\hline
$\llbracket a \rrbracket$ & encrypted vector of the individual bits of $a$\\
\hline
$\hat{a}$ & binary bit\\
\hline
$(a)_B^{(i)}$ & the $i^{th}$ bit of binary number $a$\\
\hline
\end{tabular}}
\end{table}%

\vspace{-1em}

\subsection{Skyline Definitions}

\begin{definition}(\textbf{Skyline}).
Given a dataset $P=\{\textbf{p}_1,...,\textbf{p}_n\}$ in $m$-dimensional space. Let $\textbf{p}_a$ and $\textbf{p}_b$ be two different points in $P$, we say $\textbf{p}_a$ dominates $\textbf{p}_b$, denoted by $\textbf{p}_a\prec \textbf{p}_b$, if for all $j$, $\textbf{p}_a[j]\leq \textbf{p}_b[j]$, and for at least one $j,~\textbf{p}_a[j]< \textbf{p}_b[j]$, where $\textbf{p}_i[j]$ is the $j^{th}$ dimension of $\textbf{p}_i$ and $1\leq j\leq m$. The skyline points are those points that are not dominated by any other point in $P$.
\end{definition}

\begin{definition}(\textbf{Dynamic Skyline Query}) \cite{DBLP:conf/vldb/DellisS07}.\label{def:dyskyline}
Given a dataset $P=\{\textbf{p}_1,...,\textbf{p}_n\}$ and a query point $\textbf{q}$ in $m$-dimensional space. Let $\textbf{p}_a$ and $\textbf{p}_b$ be two different points in $P$, we say $\textbf{p}_a$ dynamically dominates $\textbf{p}_b$ with regard to the query point $\textbf{q}$, denoted by $\textbf{p}_a\prec \textbf{p}_b$, if for all $j$, $|\textbf{p}_a[j]-\textbf{q}[j]|\leq |\textbf{p}_b[j]-\textbf{q}[j]|$, and for at least one $j,~|\textbf{p}_a[j]-\textbf{q}[j]|<|\textbf{p}_b[j]-\textbf{q}[j]|$, where $\textbf{p}_i[j]$ is the $j^{th}$ dimension of $\textbf{p}_i$ and $1\leq j\leq m$. The skyline points are those points that are not dynamically dominated by any other point in $P$.
\end{definition}
The traditional skyline definition is a special case of dynamic skyline query in which the query point is the origin. On the other hand, dynamic skyline query is equivalent to traditional skyline computation if we map the points to a new space with the query point $\textbf{q}$ as the origin and the absolute distances to $\textbf{q}$ as mapping functions. So the protocols we will present in the paper also work for traditional skyline computation (without an explicit query point).

\begin{example}
Consider Table \ref{tab:heartD} and Figure \ref{fig:initEG} as a running example. Given data points $\textbf{p}_1$ to $\textbf{p}_4$ and query point $\textbf{q}$, the mapped data points are computed as $\textbf{t}_i[j]=|\textbf{p}_i[j]-\textbf{q}[j]|+\textbf{q}[j]$. We see that $\textbf{t}_1, \textbf{t}_2$ are the skyline in the mapped space, and $\textbf{p}_1, \textbf{p}_2$ are the skyline with respect to query $\textbf{q}$ in the original space.
\end{example}

\subsection{Skyline Computation}

Skyline computation has been extensively studied as we discussed in Section 2. We illustrate an iterative skyline computation algorithm (Algorithm \ref{Alg:skycom}) which will be used as the basis of our secure skyline protocol. We note that this is not the most efficient algorithm to compute skyline for plaintext compared to the divide-and-conquer algorithm \cite{DBLP:journals/jacm/KungLP75}. We construct our secure skyline protocol based on this algorithm for two reasons: 1) the divide-and-conquer approach is less suitable if not impossible for a secure implementation compared to the iterative approach, 2) the performance of the divide-and-conquer algorithm deteriorate with the ``curse of dimensionality''.

The general idea of Algorithm \ref{Alg:skycom} is to first map the data points to the new space with the query point as origin (Lines 1-3). Given the new data points, it computes the sum of all attributes for each tuple $S(\textbf{t}_i)$ (Line 6) and chooses the tuple $\textbf{t}_{min}$ with smallest $S(\textbf{t}_i)$ as a skyline because no other tuples can dominate it. It then deletes those tuples dominated by $\textbf{t}_{min}$. The algorithm repeats this process for the remaining tuples until an empty dataset $T$ is reached.

\begin{algorithm}[h]\small  \caption{Skyline Computation.}\label{Alg:skycom}
\SetKwInOut{Input}{input}\SetKwInOut{Output}{output}

\Input{A dataset $P$ and a query $\textbf{q}$.}
\Output{Skyline of $P$.}

\For{$i=1$ to $n$}{
\For{$j=1$ to $m$}{
$\textbf{t}_i[j]=|\textbf{p}_i[j]-\textbf{q}[j]|$\;
}}

\While{the dataset $T$ is not empty}{
\For{$i=1$ to size of dataset $T$}{
$S(\textbf{t}_i)=\sum_{j=1}^m \textbf{t}_{i}[j]$\;
choose the tuple $\textbf{t}_{min}$ with smallest $S(\textbf{t}_i)$ as a skyline\;
add corresponding tuple $\textbf{p}_{min}$ to the skyline pool\;
delete those tuples dominated by $\textbf{t}_{min}$ from $T$\;
delete tuple $\textbf{t}_{min}$ from $T$\;
}}
\Return{skyline pool}\;
\end{algorithm}

\vspace{-1em}
\begin{example}
Given the mapped data points $\textbf{t}_1, ..., \textbf{t}_4$, we begin by computing the attribute sum for each tuple as $S(\textbf{t}_1)=16$, $S(\textbf{t}_2)=7$, $S(\textbf{t}_3)=9$, and $S(\textbf{t}_4)=19$. We choose the tuple with smallest $S(\textbf{t}_i)$, i.e., $\textbf{t}_2$, as a skyline tuple, delete $\textbf{t}_2$ from dataset $T$ and add $\textbf{p}_2$ to the skyline pool. We then delete tuples $\textbf{t}_3$ and $\textbf{t}_4$ from $T$ because they are dominated by $\textbf{t}_2$. Now, there is only $\textbf{t}_1$ in $T$. We add $\textbf{p}_1$ to the skyline pool. After deleting $\textbf{t}_1$ from $T$, $T$ is empty and the algorithm terminates. $\textbf{p}_1$ and $\textbf{p}_2$ in the skyline pool are returned as the query result.
\end{example}

\subsection{Problem Setting}

We now describe our problem setting for secure skyline queries over encrypted data. Consider a data owner (e.g., hospital, CDC) with a dataset $P$. Before outsourcing the data, the data owner encrypts each attribute of each record $\textbf{p}_i[j]$ using a semantically secure public-key cryptosystem. Fully homomorphic encryption schemes ensure strong security while enabling arbitrary computations on the encrypted data. However, the computation cost is prohibitive in practice. Partially homomorphic encryption is much more efficient but only provides partially (either additive or multiplicative) homomorphic properties.  Among them, we chose Paillier \cite{DBLP:conf/eurocrypt/Paillier99} mainly due to its additive homomorphic properties as we employ significantly more additions than multiplications in our protocol. Furthermore, we can also utilize its homomorphic multiplication between ciphertext and plaintext. We use $pk$ and $sk$ to denote the public key and private key, respectively. Data owner sends $E_{pk}(\textbf{p}_i[j])$ for $i=1,..., n$ and $j=1,..., m$ to cloud server $\mathcal{C}_1$.

\begin{figure}[htb]
 \centering
 \includegraphics[width=0.48\textwidth]{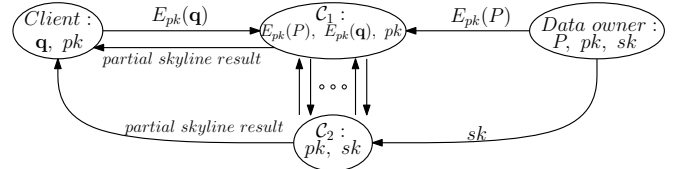}
 \vspace{-1em}
 \caption{Overview of protocol setting.}
 \label{fig:probStat}
\end{figure}

Consider an authorized client (e.g., physician) who wishes to query the skyline tuples corresponding to query tuple $\textbf{q}=( \textbf{q}[1],...,\textbf{q}[m])$. In order to protect the sensitive query tuple, the client uses the same public key $pk$ to encrypt the query tuple and sends $E_{pk}(\textbf{q})=( E_{pk}(\textbf{q}[1]),...,E_{pk}(\textbf{q}[m]))$ to cloud server $\mathcal{C}_1$.

Our goal is to enable the cloud server to compute and return the skyline to the client without learning any information about the data and the query.  In addition to guaranteeing the correctness of the result and the efficiency of the computation, the computation should require no or minimal interaction from the client or the data owner for practicality.  To achieve this, we assume there is an additional non-colluding cloud server, $\mathcal{C}_2$, which will hold the private key $sk$ shared by the data owner and assist with the computation.  This way, the data owner does not need to participate in any computation. The client also does not need to participate in any computation except combining the partial result from $\mathcal{C}_1$ and $\mathcal{C}_2$ for final result.  An overview of the protocol setting is shown in Figure \ref{fig:probStat}.

\subsection{Security Model}\label{subsec:securityModel}

\partitle{Adversary Model} We adopt the \emph{semi-honest} adversary model in our study.  In any multi-party computation setting, a \emph{semi-honest} party correctly follows the protocol specification, yet attempts to learn additional information by analyzing the transcript of messages received during the execution. By semi-honest model, this work implicitly assumes that the two cloud servers do not collude.

There are two main reasons to adopt the semi-honest adversary model in our study. First, developing protocols under the semi-honest setting is an important first step towards constructing protocols with stronger security guarantees \cite{DBLP:conf/uss/HuangEKM11}.  
Using zero-knowledge proofs \cite{DBLP:journals/joc/FeigeFS88}, these protocols can be transformed into secure protocols  under the malicious model.
Second, the semi-honest model is realistic in current cloud market. $\mathcal{C}_1$ and $\mathcal{C}_2$ are assumed to be two cloud servers, which are legitimate, well-known companies (e.g., Amazon, Google, and Microsoft). A collusion between them is highly unlikely. Therefore, following the work done in \cite{DBLP:conf/icde/ElmehdwiSJ14, DBLP:conf/icde/LiuZLLZZ15, DBLP:conf/edbt/ZhuMK14}, we also adopt the semi-honest adversary model for this paper. Please see Security Definition in the Semi-honest Model and Paillier Cryptosystem in the appendix.

\partitle{Desired Privacy Properties}
Our security goal is to protect the data and the query as well as the query result from the cloud servers.  We summarize the desired privacy properties below.  After the execution of the entire protocol, the following should be achieved.

\begin{itemize}
\item \textbf{Data Privacy.} Cloud servers $\mathcal{C}_1$ and $\mathcal{C}_2$ know nothing about the exact data except the size pattern, the client knows nothing about the dataset except the skyline query result.
\item \textbf{Data Pattern Privacy.} Cloud servers $\mathcal{C}_1$ and $\mathcal{C}_2$ know nothing about the data patterns (indirect data knowledge) due to intermediate result, e.g., which tuple dominates which other tuple.
\item \textbf{Query Privacy.} Data owner, cloud servers $\mathcal{C}_1$ and $\mathcal{C}_2$ know nothing about the query tuple $\textbf{q}$.
\item \textbf{Result Privacy.} Cloud servers $\mathcal{C}_1$ and $\mathcal{C}_2$ know nothing about the query result, e.g., which tuples are in the skyline result.
\end{itemize}

\section{Basic Security Subprotocols}\label{sec:primitives}
In this section, we present a set of secure subprotocols for computing basic functions on encrypted data that will be used to construct our secure skyline query protocol.  All protocols assume a two-party setting, namely, $\mathcal{C}_1$ with encrypted input and $\mathcal{C}_2$ with the private key $sk$ as shown in Figure \ref{fig:probStat}. The goal is for $\mathcal{C}_1$ to obtain an encrypted result of a function on the input without disclosing the original input to either $\mathcal{C}_1$ or $\mathcal{C}_2$. We note that this is different from the traditional two-party secure computation setting with techniques such as garbled circuits \cite{DBLP:conf/uss/HuangEKM11} where each party holds a private input and they wish to compute a function of the inputs. For each function, we describe the input and output, present our proposed protocol or provide a reference if existing solutions are available. Due to limited space, we omit the security proof which can be derived by the simulation and composition theorem in a straightforward way. Please see Secure Multiplication (SM), Secure Bit Decomposition (SBD), and Secure Boolean Operations in the appendix.

\subsection{Secure Minimum and Secure Comparison}

Secure minimum protocol and secure comparison protocol have been extensively studied in cryptography community \cite{DBLP:conf/fc/BaldimtsiO15, DBLP:conf/pet/ErkinFGKLT09, DBLP:journals/jstsp/VeugenBHE15} and database community \cite{ DBLP:conf/icde/ElmehdwiSJ14, DBLP:conf/edbt/ZhuMK14}. Secure comparison protocol can be easily adapted to secure minimum protocol, and vice versa. For example, if we set $E_{pk}(out)$ as the result of secure comparison $E_{pk}(Bool(a\leq b))$ known by cloud server $\mathcal{C}_1$ (it will be $E_{pk}(1)$ when $a\leq b$ and $E_{pk}(0)$ when $a>b$), $\mathcal{C}_1$ can get $E_{pk}(min(a,b))$ by computing $E_{pk}(a*out+b*\neg out)$.

We analyzed the existing protocols and observed that both secure minimum (SMIN) algorithms \cite{ DBLP:conf/icde/ElmehdwiSJ14,DBLP:conf/edbt/ZhuMK14} from database community for selecting a minimum have a security weakness, i.e., $\mathcal{C}_2$ can determine whether the two numbers are equal to each other. We point out the security weakness in the appendix.

Therefore, we adapted the secure minimum/comparison protocols \cite{DBLP:journals/jstsp/VeugenBHE15} from cryptography community in this paper. The basic idea of those protocols is that for any two $l$ bit numbers $a$ and $b$, the most significant bit ($z_l$) of $z=2^l+a-b$ indicates the relationship between $a$ and $b$, i.e., $z_l=0\Leftrightarrow a<b$. We list the secure minimum/comparison protocols we used in this paper below.

\partitle{Secure Less Than or Equal (SLEQ)}
Assume a cloud server $\mathcal{C}_1$ with encrypted input $E_{pk}(a)$ and $E_{pk}(b)$, and a cloud server $\mathcal{C}_2$ with the private key $sk$, where $a$ and $b$ are two numbers not known to $\mathcal{C}_1$ and $\mathcal{C}_2$. The goal of the SLEQ protocol is to securely compute the encrypted boolean output $E_{pk}(Bool(a\leq b))$, such that only $\mathcal{C}_1$ knows $E_{pk}(Bool(a\leq b))$ and no information related to $a$ and $b$ is revealed to $\mathcal{C}_1$ or $\mathcal{C}_2$.

\partitle{Secure Equal (SEQ)}
Assume a cloud server $\mathcal{C}_1$ with encrypted input $E_{pk}(a)$ and $E_{pk}(b)$, and a cloud server $\mathcal{C}_2$ with the private key $sk$, where $a$ and $b$ are two numbers not known to $\mathcal{C}_1$ and $\mathcal{C}_2$. The goal of the SEQ protocol is to securely compute the encrypted boolean output $E_{pk}(Bool(a==b))$, such that only $\mathcal{C}_1$ knows $E_{pk}(Bool(a==b))$ and no information related to $Bool(a==b)$ is revealed to $\mathcal{C}_1$ or $\mathcal{C}_2$.

\partitle{Secure Less (SLESS)}
Assume a cloud server $\mathcal{C}_1$ with encrypted input $E_{pk}(a)$ and $E_{pk}(b)$, and a cloud server $\mathcal{C}_2$ with the private key $sk$, where $a$ and $b$ are two numbers not known to $\mathcal{C}_1$ and $\mathcal{C}_2$. The goal of the SLESS protocol is to securely compute the encrypted boolean output $E_{pk}(Bool(a<b))$, such that only $\mathcal{C}_1$ knows $E_{pk}(Bool(a<b))$ and no information related to $Bool(a<b)$ is revealed to $\mathcal{C}_1$ or $\mathcal{C}_2$. This can be simply implemented by conjunction from the output of SEQ and SLEQ.

\partitle{Secure Minimum (SMIN)}
 Assume a cloud server $\mathcal{C}_1$ with encrypted input $E_{pk}(a)$ and $E_{pk}(b)$, and a cloud server $\mathcal{C}_2$ with the private key $sk$, where $a$ and $b$ are two numbers not known to both parties. The goal of the SMIN protocol is to securely compute encrypted minimum value of $a$ and $b$, $E_{pk}(min(a,b))$, such that only $\mathcal{C}_1$ knows $E_{pk}(min(a,b))$ and no information related to $a$ and $b$ is revealed to $\mathcal{C}_1$ or $\mathcal{C}_2$. Benefiting from the probabilistic property of Paillier, the ciphertext of $min(a,b)$, i.e., $E_{pk}(min(a,b))$ is different from the ciphertext of $a$, $b$, i.e., $E_{pk}(a)$, $E_{pk}(b)$. Therefore, $\mathcal{C}_1$ does not know which of $a$ or $b$ is $min(a,b)$. In general, assume $\mathcal{C}_1$ has $n$ encrypted values, the goal of SMIN protocol is to securely compute encrypted minimum of the $n$ values.

\section{Secure Dominance Protocol}\label{sec:secDom}

The key to compute skyline is to compute dominance relationship between two tuples. Assume a cloud server $\mathcal{C}_1$ with encrypted tuples $\textbf{a}=( \textbf{a}[1],...,\textbf{a}[m])$, $\textbf{b}=( \textbf{b}[1],...,\textbf{b}[m])$ and a cloud server $\mathcal{C}_2$ with the private key $sk$, where $\textbf{a}$ and $\textbf{b}$ are not known to both parties. The goal of the secure dominance (SDOM) protocol is to securely compute $E_{pk}(Bool(\textbf{a}\prec \textbf{b}))$ such that only $\mathcal{C}_1$ knows $E_{pk}(1)$ if $\textbf{a}\prec \textbf{b}$, otherwise, $E_{pk}(0)$.

\partitle{Protocol Design} Given any two tuples $\textbf{a}=( \textbf{a}[1],...,\textbf{a}[m])$ and $\textbf{b}=( \textbf{b}[1],...,\textbf{b}[m])$, recall the definition of skyline, we say $\textbf{a}\prec \textbf{b}$ if for all $j,~\textbf{a}[j]\leq \textbf{b}[j]$ and for at least one $j$, $\textbf{a}[j]<\textbf{b}[j]~ (1\leq j\leq m)$.
If for all $j,~\textbf{a}[j]\leq \textbf{b}[j]$, we have either $\textbf{a}=\textbf{b}$ or $\textbf{a}\prec \textbf{b}$. We refer to this case as $\textbf{a}\preceq \textbf{b}$. The basic idea of secure dominance protocol is to first determine whether $\textbf{a}\preceq \textbf{b}$, and then determine whether $\textbf{a}=\textbf{b}$.

The detailed protocol is shown in Algorithm \ref{Alg:sdom}. For each attribute, $\mathcal{C}_1$ and $\mathcal{C}_2$ cooperatively use the secure less than or equal (SLEQ) protocol to compute $E_{pk}(Bool(\textbf{a}[j]\leq \textbf{b}[j]))$. And then $\mathcal{C}_1$ and $\mathcal{C}_2$ cooperatively use SAND to compute $\Phi=\delta_1\wedge, ...,\wedge \delta_m$. If $\Phi=E_{pk}(1)$, it means $\textbf{a}\preceq \textbf{b}$, otherwise, $\textbf{a}\npreceq \textbf{b}$. We note that, the dominance relationship information $\Phi$ is known only to $\mathcal{C}_1$ in ciphertext. Therefore, both $\mathcal{C}_1$ and $\mathcal{C}_2$ do not know any information about whether $\textbf{a}\preceq \textbf{b}$.

\begin{algorithm}[h] \small \caption{Secure Dominance Protocol.}\label{Alg:sdom}
\SetKwInOut{Input}{input}\SetKwInOut{Output}{output}

\Input{$\mathcal{C}_1$ has $E_{pk}(\textbf{a}),E_{pk}(\textbf{b})$ and $\mathcal{C}_2$ has $sk$.}
\Output{$\mathcal{C}_1$ gets $E_{pk}(1)$ if $\textbf{a}\prec \textbf{b}$, otherwise, $\mathcal{C}_1$ gets $E_{pk}(0)$.}

$\mathcal{C}_1$ and $\mathcal{C}_2$:\\
\For{$j=1$ to $m$}{
$\mathcal{C}_1$ gets $\delta_j=E_{pk}(Bool(\textbf{a}[j]\leq \textbf{b}[j]))$ by SLEQ\;
}
use SAND to compute $\Phi=\delta_1\wedge ...,\wedge \delta_m$\;
$\mathcal{C}_1$:\\
compute $\alpha=E_{pk}(\textbf{a}[1])\times ,..., \times E_{pk}(\textbf{a}[m])$\;
compute $\beta=E_{pk}(\textbf{b}[1])\times ,..., \times E_{pk}(\textbf{b}[m])$\;

$\mathcal{C}_1$ and $\mathcal{C}_2$:\\
$\mathcal{C}_1$ gets $\sigma=E_{pk}(Bool(\alpha<\beta))$ by employing SLESS\;
$\mathcal{C}_1$ gets $\Psi=\sigma \wedge \Phi$ as the final dominance relationship using  SAND\;
\end{algorithm}

Next, we need to determine if $\textbf{a}\neq \textbf{b}$. Only if $\textbf{a}\neq \textbf{b}$, then $\textbf{a}\prec \textbf{b}$. One naive way is to employ SEQ protocol for each pair of attribute and then take the conjunction of the output.  We propose a more efficient way which is to check whether $S(\textbf{a})< S(\textbf{b})$, where $S(\textbf{a})$ is the attribute sum of tuple $\textbf{a}$. If $S(\textbf{a}) < S(\textbf{b})$, then it is impossible that $\textbf{a}=\textbf{b}$. As the algorithm shows,
$\mathcal{C}_1$ computes the sum of all attributes
$\alpha=E_{pk}(\textbf{a}[1]+...+\textbf{a}[m])$ and $\beta=E_{pk}(\textbf{b}[1]+...+\textbf{b}[m])$ based on the additive homomorphic property.
Then $\mathcal{C}_1$ and $\mathcal{C}_2$ cooperatively use SLESS protocol to compute $\sigma=E_{pk}(Bool(\alpha<\beta))$. Finally, $\mathcal{C}_1$ and $\mathcal{C}_2$ cooperatively use SAND protocol to compute the final dominance relationship $\Psi=\sigma \wedge \Phi$ which is only known to $\mathcal{C}_1$ in ciphertext. $\Psi=E_{pk}(1)$ means $\textbf{a}\prec \textbf{b}$, otherwise, $\textbf{a}\nprec \textbf{b}$.

\partitle{Security Analysis}
Based on the composition theorem (Theorem \ref{def:comp}), the security of secure dominance protocol relies on the security of SLEQ, SLESS, and SAND, which have been shown in existing works. 

\partitle{Complexity Analysis}
To determine $\textbf{a}\preceq \textbf{b}$, Algorithm \ref{Alg:sdom} requires $O(m)$ encryptions and decryptions. Then to determine if $\textbf{a} = \textbf{b}$, Algorithm \ref{Alg:sdom} requires $O(1)$ encryptions and decryptions. Therefore, our secure dominance protocol requires $O(m)$ encryptions and decryptions in total.

\section{Secure Skyline Protocol}\label{sec:SSkylinePro}

In this section, we first propose a basic secure skyline protocol and show why such a simple solution is not secure. Then we propose a fully secure skyline protocol. Both protocols are constructed by using the security primitives discussed in Section \ref{sec:primitives} and the secure dominance protocol in Section \ref{sec:secDom}.

As mentioned in Algorithm \ref{Alg:skycom}, given a skyline query $\textbf{q}$, it is equivalent to compute the skyline in a transformed space with the query point $\textbf{q}$ as the origin and the absolute distances to $\textbf{q}$ as mapping functions.  Hence we first show a preprocessing step in Algorithm \ref{Alg:preproc} which maps the dataset to the new space. Since the skyline only depends on the order of the attribute values, we use $(\textbf{p}_i[j]-\textbf{q}[j])^2$ which is easier to compute than $|\textbf{p}_i[j]-\textbf{q}[j]|$ as the mapping function\footnote{We use $|\textbf{p}_i[j]-\textbf{q}[j]|$ in our running example for simplicity.}.  After Algorithm \ref{Alg:preproc}, $\mathcal{C}_1$ has the encrypted dataset $E_{pk}(P)$ and $E_{pk}(T)$, $\mathcal{C}_2$ has the private key $sk$. The goal is to securely compute the skyline by $\mathcal{C}_1$ and $\mathcal{C}_2$ without participation of data owner and the client.

\begin{algorithm}[h]\small \caption{Preprocessing.}\label{Alg:preproc}
\SetKwInOut{Input}{input}\SetKwInOut{Output}{output}

\Input{$\mathcal{C}_1$ has $E_{pk}(P)$, $\mathcal{C}_2$ has $sk$, and the client has $\textbf{q}$.}
\Output{$\mathcal{C}_1$ obtains the new encrypted dataset $E_{pk}(T)$.}

Client:\\
send $( E_{pk}(-\textbf{q}[1]),...,E_{pk}(-\textbf{q}[m]))$ to $\mathcal{C}_1$\;

$\mathcal{C}_1$:\\
\For{$i=1$ to $n$}{
\For{$j=1$ to $m$}{
$E_{pk}(temp_i[j])=E_{pk}(\textbf{p}_i[j]-\textbf{q}[j])=E_{pk}(\textbf{p}_i[j])\times E_{pk}(-\textbf{q}[j]) \mod N^2$\;
      }}

$\mathcal{C}_1$ and $\mathcal{C}_2$:\\
use SM protocol to compute $E_{pk}(T)=(E_{pk}(\textbf{t}_1),...,E_{pk}(\textbf{t}_n))$ only known by $\mathcal{C}_1$, where $E_{pk}(\textbf{t}_i)=(E_{pk}(\textbf{t}_i[1]),...,E_{pk}(\textbf{t}_i[m]))$ and $E_{pk}(\textbf{t}_i[j])=E_{pk}(temp_i[j])\times E_{pk}(temp_i[j])$\;

\end{algorithm}

\vspace{-1em}
\subsection{Basic Protocol}\label{subsec:basicProtocol}

We first illustrate a straw-man protocol which is straightforward but not fully secure (as shown in Algorithm \ref{Alg:basicProto}). The idea is to implement each of the steps in Algorithm \ref{Alg:skycom} using the primitive secure protocols.  $\mathcal{C}_1$ first determines the terminal condition, if there is no tuple exists in dataset $E_{pk}(T)$, the protocol ends, otherwise, the protocol proceeds as follows.

\partitle{Compute minimum attribute sum} $\mathcal{C}_1$ first computes the sum of $E_{pk}(\textbf{t}_i[j])$ for $1\leq j\leq m$, denoted as $E_{pk}(S(\textbf{t}_i))$, for each tuple $\textbf{t}_i$.  Then $\mathcal{C}_1$ and $\mathcal{C}_2$ uses SMIN protocol such that $\mathcal{C}_1$ obtains $E_{pk}(S(\textbf{t}_{min})) $.

\partitle{Select the skyline with minimum attribute sum} The challenge now is we need to select the tuple $E_{pk}(\textbf{t}_{min})$ with the smallest $E_{pk}(S(\textbf{t}_i))$ as a skyline tuple.
In order to do this, a naive way is for $\mathcal{C}_1$ to compute $E_{pk}(S(\textbf{t}_i)-S(\textbf{t}_{min}))$ for all tuples and then send them to $\mathcal{C}_2$.  $\mathcal{C}_2$ can decrypt them and determine which one is equal to 0 and return the index to $\mathcal{C}_1$.
$\mathcal{C}_1$ then adds the tuple $E_{pk}(\textbf{p}_{min})$ to skyline pool.

\partitle{Eliminate dominated tuples} Once the skyline tuple is selected, $\mathcal{C}_1$ and $\mathcal{C}_2$ cooperatively use SDOM protocol to determine the dominance relationship between $E_{pk}(\textbf{t}_{min})$ and other tuples. In order to delete those tuples that are dominated by $E_{pk}(\textbf{t}_{min})$, a naive way is for $\mathcal{C}_1$ to send the encrypted dominance output to $\mathcal{C}_2$, who can decrypt it and send back the indexes of the tuples who are dominated to $\mathcal{C}_2$.  $\mathcal{C}_1$ can delete those tuples dominated by $E_{pk}(\textbf{t}_{min})$ and the tuple $E_{pk}(\textbf{t}_{min})$ from $E_{pk}(T)$.  The algorithm continues until there is no tuples left.

\partitle{Return skyline results to client}
Once $\mathcal{C}_1$ has the encrypted skyline result, it can directly send them to the client if the client has the private key.  However, in our setting, the client does not have the private key for better security.  Lines 25 to 39 in Algorithm \ref{Alg:basicProto} illustrate how the client obliviously obtains the final skyline query result with the help of $\mathcal{C}_1$ and $\mathcal{C}_2$, at the same time, $\mathcal{C}_1$ and $\mathcal{C}_2$ know nothing about the final result. Consider the skyline tuples $E_{pk}(\textbf{p}_1),...,E_{pk}(\textbf{p}_k)$ in skyline pool, where $k$ is the number of skyline. The idea is for $\mathcal{C}_1$ to add a random noise $r_i[j]$ to each $\textbf{p}_i[j]$ in ciphertext and then sends the encrypted randomized values $\alpha_i[j]$ to $\mathcal{C}_2$. $\mathcal{C}_1$ also sends the noise $r_i[j]$ to client. At the same time, $\mathcal{C}_2$ decrypts the randomized values $\alpha_i[j]$ and sends the result $r'_i[j]$ to client. Client receives the random noise $r_i[j]$ from $\mathcal{C}_1$ and randomized values of the skyline points $\alpha_i[j]$ from $\mathcal{C}_2$, and removes the noise by computing $\textbf{p}_i[j]=r'_i[j]-r_i[j]$ for $i=1,...,k$ and $j=1,...,m$ as the final result.

\begin{algorithm}[h] \small \caption{Basic Secure Skyline Protocol.}\label{Alg:basicProto}
\SetKwInOut{Input}{input}\SetKwInOut{Output}{output}

\Input{$\mathcal{C}_1$ has $E_{pk}(P),E_{pk}(T)$ and $\mathcal{C}_2$ has $sk$.}
\Output{client knows the skyline query result.}

\textbf{Compute minimum attribute sum\;}
$\mathcal{C}_1$:\\
\If{there is no tuple in $E_{pk}(T)$}{
break\;}

\For{$i=1$ to $n$}{
$E_{pk}(S(\textbf{t}_i))=E_{pk}(\textbf{t}_i[1])\times ... \times E_{pk}(\textbf{t}_i[m]) \mod N^2$\;}

$\mathcal{C}_1$ and $\mathcal{C}_2$:\\
$ E_{pk}(S(\textbf{t}_{min})) =SMIN(E_{pk}(S(\textbf{t}_1)) ,..., E_{pk}(S(\textbf{t}_n)) )$\;

\textbf{Select the skyline with minimum attribute sum\;}
$\mathcal{C}_1$:\\
\For{$i=1$ to $n$}{
$\alpha_i=E_{pk}(S(\textbf{t}_{min}))^{N-1} \times E_{pk}(S(\textbf{t}_i)) \mod N^2$\;
$\alpha_i'=\alpha_i^{r_i} \mod N^2$, where $r_i\in \mathbb{Z}_N^{+} $\;
}
send $\alpha'$ to $\mathcal{C}_2$\;
$\mathcal{C}_2$:\\
decrypt $\alpha'$ and tell $\mathcal{C}_1$ which one equals to $0$\;

$\mathcal{C}_1$:\\
add the corresponding $E_{pk}(\textbf{p}_{min})$ to the skyline pool\;
\textbf{Eliminate dominated tuples\;}
$\mathcal{C}_1$ and $\mathcal{C}_2$:\\
use SDOM protocol to determine the dominance relationship between $E_{pk}(\textbf{t}_{min})$ and other tuples\;
delete those tuples dominated by $E_{pk}(\textbf{t}_{min})$ and $E_{pk}(\textbf{t}_{min})$\;

GOTO Line $1$\;
\textbf{Return skyline results to client\;}
$\mathcal{C}_1$:\\
\For{$i=1$ to $k$}{
\For{$j=1$ to $m$}{
$\alpha_i[j]=E_{pk}(\textbf{p}_i[j])\times E_{pk}(r_i[j]) \mod N^2$, where $r_i[j]\in \mathbb{Z}_N^{+}$\;}}
send $\alpha_i[j]$ to $\mathcal{C}_2$ and $r_i[j]$ to client for all $i=1,...,k; j=1,...,m$\;

$\mathcal{C}_2$:\\
\For{$i=1$ to $k$}{
\For{$j=1$ to $m$}{
$r_i[j]'=D_{sk}(\alpha_i[j])$\;}}
send $r_i[j]'$ to client\;

Client:\\
receive $r_i[j]$ from $\mathcal{C}_1$ and $r_i[j]'$ from $\mathcal{C}_2$\;
\For{$i=1$ to $k$}{
\For{$j=1$ to $m$}{
$\textbf{p}_i[j]=r_i[j]'-r_i[j]$\;}}
\end{algorithm}

\subsection{Fully Secure Skyline Protocol}\label{subsec:fullyProtocol}

The basic protocol clearly reveals several information to $\mathcal{C}_1$ and $\mathcal{C}_2$ as follows.

\begin{itemize}
\item When selecting the skyline tuple with minimum attribute sum, $\mathcal{C}_1$ and $\mathcal{C}_2$ know which tuples are skyline points, which violates our result privacy requirement.
\item When eliminating dominated tuples, $\mathcal{C}_1$ and $\mathcal{C}_2$ know the dominance relationship among tuples with respect to the query tuple $\textbf{q}$, which violates our data pattern privacy requirement.
\end{itemize}

To address these leakage, we propose a fully secure protocol in Algorithm \ref{Alg:fullyProto}. The step to compute minimum attribute sum and return the results to the client are the same as the basic protocol.  We focus on the following steps that are designed to address the disclosures of the basic protocol.

\partitle{Select skyline with minimum attribute sum} 
Once $\mathcal{C}_1$ obtains the encrypted minimum attribute sum $E_{pk}(S(\textbf{t}_{min}))$, the challenge is how to select the tuple $E_{pk}(\textbf{t}_{min})$ with the minimum sum $E_{pk}(S(\textbf{t}_{min}))$ as a skyline tuple such that $\mathcal{C}_1$ and $\mathcal{C}_2$ know nothing about which tuple is selected. We present a protocol as shown in Algorithm \ref{Alg:findOneSky}.

\begin{algorithm}[thb] \small \caption{Fully Secure Skyline Protocol.}\label{Alg:fullyProto}
\SetKwInOut{Input}{input}\SetKwInOut{Output}{output}

\Input{$\mathcal{C}_1$ has $E_{pk}(P),E_{pk}(T)$ and $\mathcal{C}_2$ has $sk$.}
\Output{$\mathcal{C}_1$ knows the encrypted skyline $E_{pk}(\textbf{p}_{sky})$.}

\textbf{Order preserving perturbation\;}
$\mathcal{C}_1$:\\
\For{$i=1$ to $n$}{
$E_{pk}(S(\textbf{t}_i))=E_{pk}(\textbf{t}_i[1])\times ... \times E_{pk}(\textbf{t}_i[m]) \mod N^2$\;}

$\mathcal{C}_1$ and $\mathcal{C}_2$:\\
\For{$i=1$ to $n$}{
$\llbracket E_{pk}(S(\textbf{t}_i)) \rrbracket=SBD(E_{pk}(S(\textbf{t}_i)))$\;}

$\mathcal{C}_1$:\\
\For{$i=1$ to $n$}{
$\llbracket E_{pk}(S(\textbf{t}_i)) \rrbracket = \langle E_{pk}((S(\textbf{t}_i))_B^{(1)}), ..., E_{pk}((S(\textbf{t}_i))_B^{(l)}),$ $E_{pk}((S(\textbf{t}_i))_B^{(l+1)}), ..., E_{pk}((S(\textbf{t}_i))_B^{(l+\lceil \log n\rceil)})\rangle$, where $(S(\textbf{t}_i))_B^{(l+1)},..., (S(\textbf{t}_i))_B^{(l+\lceil \log n\rceil)}$ is the binary representation of an exclusive vale of $[0,n-1]$\;
$E_{pk}(S(\textbf{t}_{i}))=\prod_{\gamma=1}^{l} E_{pk}((S(\textbf{t}_{i}))_B^{(\gamma)})^{2^{l-\gamma}} \mod N^2$\;}

$\mathcal{C}_1$ and $\mathcal{C}_2$:\\
$E_{pk}(S(\textbf{t}_{min})) =SMIN( E_{pk}(S(\textbf{t}_1)) ,..., E_{pk}(S(\textbf{t}_n))$\;

$\mathcal{C}_1$:\\
$\lambda=(E_{pk}(S(\textbf{t}_{min}))\times E_{pk}(MAX)^{-1})^{r} \mod N^2$, where $r_i\in  \mathbb{Z}_N^{+}$\;
send $\lambda$ to $\mathcal{C}_2$\;

$\mathcal{C}_2$:\\
\If{$D_{sk}(\lambda)=0$}{
break\;}

\textbf{Select skyline with minimum attribute sum\;}

$(E_{pk}(\textbf{p}_{sky}),E_{pk}(\textbf{t}_{sky}))=$FindOneSkyline $(E_{pk}(P),E_{pk}(T),E_{pk}(S(\textbf{t}_i)),E_{pk}(S(\textbf{t}_{min})))$ (Algorithm \ref{Alg:findOneSky})\;

\textbf{Eliminate dominated tuples\;}
$\mathcal{C}_1$ and $\mathcal{C}_2$:\\
\For{$i=1$ to $n$}{
\For{$\gamma=1$ to $l$}{
$E_{pk}((S(\textbf{t}_i))_B^{(\gamma)})=SOR(V_i,E_{pk}((S(\textbf{t}_i))_B^{(\gamma)}))$\;
}}

$\mathcal{C}_1$:\\
\For{$i=1$ to $n$}{
$E_{pk}(S(\textbf{t}_i))=\prod_{\gamma=1}^{l}E_{pk}((S(\textbf{t}_i))_B^{(\gamma)})^{2^{l-\gamma}} \mod N^2$\;
}

$\mathcal{C}_1$ and $\mathcal{C}_2$:\\
\For{$i=1$ to $n$}{
$V_i=SDOM(E_{pk}(\textbf{t}_{sky}),E_{pk}(\textbf{t}_i))$\;
}
Lines 23-32\;

GOTO Line 1\;

\end{algorithm}

We first need to determine which $S(\textbf{t}_i)$ is equal to $S(\textbf{t}_{min})$. Note that this can not be achieved by the SMIN protocol which only selects the minimum value. Here we propose an efficient way, exploiting the fact that it is okay for $\mathcal{C}_2$ to know there is one equal case (since we are selecting one skyline tuple) as long as it does not know which one.  $\mathcal{C}_1$ first computes $\alpha_i'=E_{pk}((S(\textbf{t}_i)-S(\textbf{t}_{min}))\times r_i)$, and then sends a permuted list $\beta=\pi(\alpha')$ to $\mathcal{C}_2$ based on a random permutation sequence $\pi$. The permutation hides which sum is equal to the minimum from $\mathcal{C}_2$ while the uniformly random noise $r_i$ masks the difference between each sum and the minimum sum. Note that $\alpha_i'$ is uniformly random in $\mathbb{Z}_N^{+}$ except when $S(\textbf{t}_i)-S(\textbf{t}_{min})=0$, in which case $\alpha_i'=0$. $\mathcal{C}_1$ decrypts $\beta_i$, if it is 0, it means tuple $i$ has smallest $E_{pk}(S(\textbf{t}_i))$. Therefore, $\mathcal{C}_2$ sends $E_{pk}(1)$ to $\mathcal{C}_1$, otherwise, sends $E_{pk}(0)$.

After receiving the encrypted permuted bit vector $U$ as the equality result, $\mathcal{C}_1$ applies a reverse permutation, and obtains an encrypted bit vector $V$, where one tuple has bit 1 suggesting it has the minimum sum. In order to obtain the attribute values of this tuple, $\mathcal{C}_1$ and $\mathcal{C}_2$ employ SM protocol to compute encrypted product of the bit vector and the attribute values, $E_{pk}(\textbf{t}_i[j]')$ and $E_{pk}(\textbf{p}_i[j]')$.  Since all other tuples except the one with the minimum sum will be 0, we can sum all $E_{pk}(\textbf{t}_i[j]')$ and $E_{pk}(\textbf{p}_i[j]')$ on each attribute and  $\mathcal{C}_1$ can obtain the attribute values corresponding to the skyline tuple. 

\begin{algorithm}[h] \small \caption{Find One Skyline.}\label{Alg:findOneSky}
\SetKwInOut{Input}{input}\SetKwInOut{Output}{output}

\Input{$\mathcal{C}_1$ has encrypted dataset $E_{pk}(P)$, $E_{pk}(T)$, $E_{pk}(S(\textbf{t}_i))$, and $E_{pk}(S(\textbf{t}_{min}))$, $\mathcal{C}_2$ has private key $sk$.}
\Output{$\mathcal{C}_1$ knows one encrypted skyline $E_{pk}(\textbf{p}_{sky})$ and $E_{pk}(\textbf{t}_{sky})$.}

$\mathcal{C}_1$:\\
\For{$i=1$ to $n$}{
$\alpha_i=E_{pk}(S(\textbf{t}_{min}))^{N-1} \times E_{pk}(S(\textbf{t}_i)) \mod N^2$\;
$\alpha_i'=\alpha_i^{r_i} \mod N^2$, where $r_i\in  \mathbb{Z}_N^{+}$\;
}
send $\beta=\pi(\alpha')$ to $\mathcal{C}_2$\;

$\mathcal{C}_2$:\\
receive $\beta$ from $\mathcal{C}_1$\;
\For{$i=1$ to $n$}{
$\beta_i'=D_{sk}(\beta_i)$\;
\If{$\beta_i'=0$}{
$U_i=E_{pk}(1)$\;}
\Else{
$U_i=E_{pk}(0)$\;}
}
send $U$ to $\mathcal{C}_1$\;

$\mathcal{C}_1$:\\
receive $U$ from $\mathcal{C}_2$\;
$V=\pi ^{-1}(U)$\;
\For{$i=1$ to $n$}{
\For{$j=1$ to $m$}{
$E_{pk}(\textbf{t}_i[j]')=SM(V_i,E_{pk}(\textbf{t}_i[j]))$\;
$E_{pk}(\textbf{p}_i[j]')=SM(V_i,E_{pk}(\textbf{p}_i[j]))$\;}}
\For{$j=1$ to $m$}{
$E_{pk}(\textbf{t}[j]')=\prod_{i=1}^{n}E_{pk}(\textbf{t}_i[j]') \mod N^2$\;
$E_{pk}(\textbf{p}[j]')=\prod_{i=1}^{n}E_{pk}(\textbf{p}_i[j]') \mod N^2$\;
}
add $E_{pk}(\textbf{p}_{sky})=\langle E_{pk}(\textbf{p}[1]'),...,E_{pk}(\textbf{p}[m]')\rangle$ to skyline pool\;
use $E_{pk}(\textbf{t}_{sky})=\langle E_{pk}(\textbf{t}[1]'),...,E_{pk}(\textbf{t}[m]')\rangle$ to compare with other $E_{pk}(\textbf{t}_i)$\;

\end{algorithm}

\partitle{Order preserving perturbation} We can show that Algorithm \ref{Alg:findOneSky} is secure and correctly selects the skyline tuple if there is only one minimum.  A potential issue is that multiple tuples may have the same minimum sum.  If this happens, not only is this information revealed to $\mathcal{C}_2$, but also the skyline tuple cannot be selected (computed) correctly, since the bit vector contains more than one 1 bit. To address this, we employ order-preserving perturbation which adds a set of mutually different  {\em bit sequence} to {\em a set of values} such that: 1) if the original values are equal to each other, the perturbed values are guaranteed not equal to each other, and 2) if the original values are not equal to each other, their order is preserved.  The perturbed values are then used as the input for Algorithm \ref{Alg:findOneSky}.

Concretely, given $n$ numbers in their binary representations, we add a $\lceil log n\rceil$-bit sequence to the end of each $E_{pk}(S(\textbf{t}_i))$, each represents a unique bit sequence in the range of $[0,n-1]$. This way, the perturbed values are guaranteed to be different from each other while their order is preserved since the added bits are the least significant bits. Line 10 of Algorithm \ref{Alg:fullyProto} shows this step. We note that we can multiply each sum $E_{pk}(S(\textbf{t}_i))$ by $n$ and uniquely add a value from $[0, n-1]$ to each $E_{pk}(S(\textbf{t}_i))$, hence guarantee they are not equal to each other. This will be more efficient than adding a bit sequence, however, since we will need to perform the bit decomposition later in the protocol to allow bit operators, we run decomposition by the SBD protocol for $l$ bits in the beginning of the protocol rather than $l+\lceil \log n\rceil$ bits later.

\begin{table*}[thb]\centering
\caption{Example of Algorithm 5.}\label{tab:full}\scriptsize
\vspace{-1em}
{%
\begin{tabular}{|p{0.2cm}|p{1.2cm}|p{0.4cm}|p{1.1cm}|p{0.4cm}|p{0.5cm}|p{1.9cm}|p{0.2cm}|p{0.2cm}|p{0.3cm}|p{0.2cm}|p{0.2cm}|p{1.4cm}|p{1.5cm}|p{0.5cm}|p{0.2cm}|p{0.5cm}|}
\hline
\multicolumn{9}{|c|}{$\mathcal{C}_1$:}                                                                              & \multicolumn{2}{c|}{$\mathcal{C}_2$:}     & \multicolumn{6}{c|}{$\mathcal{C}_1$:}\\
\hline
$\textbf{t}_i$ & $(\textbf{t}_i[1],\textbf{t}_i[2])$ & $S(\textbf{t}_i)$ & $\llbracket S(\textbf{t}_i)\rrbracket$ & pert. & $S(\textbf{t}_i)$ & $S(\textbf{t}_i)-S(\textbf{t}_{min})$ &$r$& $\pi$ & $\beta'$ & $U$  & $V$ & $(\textbf{t}_i[1]',\textbf{t}_i[2]')$& $(\textbf{p}_i[1]',\textbf{p}_i[2]')$ & $S(\textbf{t}_i)$ & $V$ & $S(\textbf{t}_i)$\\
\hline
$\textbf{t}_1$   &  $(1,15)$                            & $16$               & $1,0,0,0,0$                            & $1,1$         & $67$              & $67-30$                  & $3$     & $2$   & $0$     & $1$  & $0$ & $(0,0)$ & $(0,0)$                               & $67$              & $0$ & $67$\\

$\textbf{t}_2$   &  $(2,5)$                             & $7$                & $0,0,1,1,1$                            & $1,0$         & $30$              & $30-30$                  & $9$     & $1$   & $111$   & $0$  & $1$ & $(2,5)$ & $(39,120)$                               & $127$             & $0$ & $127$\\

$\textbf{t}_3$   &  $(4,5)$                             & $9$                & $0,1,0,0,1$                            & $0,1$         & $37$              & $37-30$                  & $31$    & $4$   & $92$    & $0$  & $0$ & $(0,0)$  & $(0,0)$                              & $37$              & $1$ & $127$\\

$\textbf{t}_4$   &  $(4,15)$                            & $19$               & $1,0,0,1,1$                            & $0,0$         & $76$              & $76-30$                  & $2$     & $3$   & $217$   & $0$  & $0$ & $(0,0)$ & $(0,0)$                               & $76$              & $1$ & $127$\\
\hline
\end{tabular}}
\end{table*}%

\partitle{Eliminate dominated tuples} Once the skyline tuple is selected, it can be added to the skyline pool and then used to eliminate dominated tuples.  In order to do this, $\mathcal{C}_1$ and $\mathcal{C}_2$ cooperatively use SDOM protocol to determine the dominance relationship between $E_{pk}(\textbf{t}_{min})$ and other tuples.  The challenge is then how to eliminate the dominated tuples without $\mathcal{C}_1$ and $\mathcal{C}_2$ knowing which tuples are being dominated and eliminated.  Our idea is that instead of eliminating the dominated tuples, we ``flag" them by securely setting their attribute values to the maximum domain value.  This way, they will not be selected as skyline tuples in the remaining iterations. Concretely, we can set the binary representation of their attribute sum to all 1s so that it represents the domain maximum. Since we added $\lceil \log n\rceil$ bits to  $\llbracket E_{pk}(S(\textbf{t}_i)) \rrbracket$, the new $\llbracket E_{pk}(S(\textbf{t}_i)) \rrbracket$ has $l+\lceil \log n\rceil$ bits.  Therefore, the maximum value 
$MAX=2^{l+\lceil \log n\rceil}-1$. To obliviously set the attributes of only dominated tuples to $MAX$, based on the encrypted dominance output $V_i$ of the dominance protocol, $\mathcal{C}_1$ and $\mathcal{C}_2$ cooperatively employ SOR of the dominance boolean output and the bits of the $S(\textbf{t}_{i})$. This way, if the tuple is dominated, it will be set to MAX.  Otherwise, it will remain the same. If $E_{pk}(S(\textbf{t}_{min}))=E_{pk}(MAX)$, it means all the tuples are processed, i.e., flagged either as a skyline or a dominated tuple, the protocol ends.

\begin{example}
We illustrate the entire protocol through the running example shown in Table \ref{tab:full}. Please note that all column values are in encrypted form except columns $\pi$ and $\beta'$. Given the mapped data points $\textbf{t}_i$, $\mathcal{C}_1$ first computes the attribute sum $E_{pk}(S(\textbf{t}_i))$ shown in the third column. We set $l=5$, $\mathcal{C}_1$ gets the binary representation of the attribute sum $\llbracket E_{pk}(S(\textbf{t}_i)) \rrbracket$. Because $n=4$, $\mathcal{C}_1$ obliviously adds the order-preserving perturbation $\lceil \log 4\rceil=2$ bits to the end of $\llbracket E_{pk}(S(\textbf{t}_i)) \rrbracket$ respectively to get the new $E_{pk}(S(\textbf{t}_i))$ (shown in the sixth column). Then $\mathcal{C}_1$ gets $E_{pk}(S(\textbf{t}_{min}))=E_{pk}(30)$ by employing SMIN.

The protocol then turns to the subroutine Algorithm \ref{Alg:findOneSky} to select the first skyline based on the minimum attribute sum. $\mathcal{C}_1$ computes $\alpha_i=E_{pk}(S(\textbf{t}_i)-S(\textbf{t}_{min}))$. Assume the random noise vector $r=\langle 3,9,31,2\rangle$ and the permutation sequence $\pi=\langle 2, 1, 4, 3\rangle$, $\mathcal{C}_1$ sends the encrypted permuted and randomized difference vector $\beta$ to $\mathcal{C}_2$. After decrypting $\beta$, $\mathcal{C}_2$ gets $\beta'$ and then sends $U$ to $\mathcal{C}_1$. $\mathcal{C}_1$ computes $V$ by applying a reverse permutation. By employing SM with $V$, $\mathcal{C}_1$ computes $(E_{pk}(\textbf{t}_i[1]'),E_{pk}(\textbf{t}_i[2]'))$ and $(E_{pk}(\textbf{p}_i[1]'),E_{pk}(\textbf{p}_i[2]'))$. After summing all column values, $\mathcal{C}_1$ adds $E_{pk}(\textbf{p}_{sky})=(E_{pk}(39)$, $E_{pk}(120))$ to skyline pool and uses $E_{pk}(\textbf{t}_{sky})=( E_{pk}(2), E_{pk}(5))$ to eliminate dominated tuples.

The protocol now turns back to the main routine in Algorithm \ref{Alg:fullyProto} to eliminate dominated tuples. $\mathcal{C}_1$ and $\mathcal{C}_2$ use SOR with $V$ to make $E_{pk}(S(\textbf{t}_{min}))=E_{pk}(127)$ and $E_{pk}(S(\textbf{t}_{i}))=E_{pk}(S(\textbf{t}_{i}))$ for $i\neq min$. Now, only $E_{pk}(S(\textbf{t}_{min}))=E_{pk}(S(\textbf{t}_2))$ has changed to $E_{pk}(127)$  which is ``flagged" as MAX. We emphasize that $\mathcal{C}_1$ does not know this value has changed because the ciphertext of all tuples has changed. Next, $\mathcal{C}_1$ and $\mathcal{C}_2$ find the dominance relationship between $E_{pk}(\textbf{t}_{sky})$ and $E_{pk}(\textbf{t}_i)$ by SDOM protocol. $\mathcal{C}_1$ obtains the dominance vector $V$. Using same method, $\mathcal{C}_1$ flags $E_{pk}(S(\textbf{t}_3))$ and $E_{pk}(S(\textbf{t}_4))$ to $E_{pk}(127)$. The protocol continues until all are set to MAX.
\end{example}

\partitle{Security Analysis}
Based on Theorem 1, the protocol is secure if the subprotocols are secure and the intermediate results are random or pseudo-random.  We focus on the intermediate result here.  From $\mathcal{C}_1$'s view, the intermediate result includes $U$. Because $U$ is ciphertext and $\mathcal{C}_1$ does not have the secret key, $\mathcal{C}_1$ can simulate $U$ based on its input and output. 
From $\mathcal{C}_2$'s view, the intermediate result includes $\beta$. $\beta$ contains one $E_{pk}(0)$ and $m-1$ ciphertext of any positive value. After the permutation $\pi$ of $\mathcal{C}_1$, $\mathcal{C}_2$ cannot determine where is the $E_{pk}(0)$. Therefore, $\mathcal{C}_2$ can simulate $\beta$ based on its input and output. 
Hence the protocol is secure.

\partitle{Computational Complexity Analysis}
The subroutine Algorithm \ref{Alg:findOneSky} requires $O(n)$ decryptions in Line 9, $O(nm)$ encryptions and decryptions in Lines 20 and 21. Thus, Algorithm \ref{Alg:findOneSky} requires $O(nm)$ encryptions and decryptions in all. In Algorithm \ref{Alg:fullyProto}, Line 7 requires $O(nl)$ encryptions and decryptions. Line 10 requires $O(n\lceil \log n\rceil)$ encryptions. Line 12 requires $O((l+\lceil \log n\rceil)n)$ encryptions and decryptions. Line 26 requires $O(l+\lceil \log n\rceil)$ encryptions and decryptions. Line 32 requires $O(nm)$ encryptions and decryptions. Thus, this part requires $O((l+\lceil \log n\rceil)n+nm)$ encryptions and decryptions. Because this part runs $k$ times, the fully secure skyline protocol requires $O(k(l+\lceil \log n\rceil)n+knm)$ encryptions and decryptions in total.

\section{Performance Analysis and Optimizations}\label{sec:Perfor}
In this section, we illustrate two optimizations to further reduce the computation load. We first show a data partitioning optimization in Subsection \ref{subsec:dataPartition}, and then show a lazy merging optimization in Subsection \ref{subsec:lazymerge}.

\subsection{Optimization of Data Partitioning}\label{subsec:dataPartition}
As shown in the previous section, the overall run time complexity depends on the number of points ($n$), the number of skyline points ($k$), the number of decomposed bits ($l$) which is determined by the domain of the attribute values, and the number of dimensions ($m$). A straightforward way to enhance the performance is to partition the input dataset into subdatasets and then we can use a divide-and-conquer approach to avoid unnecessary computations. Furthermore, the partitioning also allows effective parallelism.

The basic idea of data partitioning is to divide the dataset into a set of initial partitions, compute the skyline in each partition, and then continuously merge the skyline result of the partitions into new partitions and compute their skyline, until all partitions are merged into the final result.  This can be implemented with either a single thread (sequentially) or multiple threads (in parallel). We describe our data partitioning scheme in Algorithm \ref{Alg:dataPartition}. Given an input dataset, the number of partitions $s$ is specified as one parameter. We will show how to calculate the optimal number of partitions in Subsection \ref{sec:partitionTheory}. We first divide the input data into $s$ partitions and compute the skyline in each partition in Line 1, and then set the state of all partitions as uncomputed in Line 2. In Line 7, the algorithm continues with uncomputed partitions or idle threads. In Line 8, if there are some uncomputed partitions and there are some idle threads, we assign one uncomputed partition to each idle thread in Line 9. In Line 11, if there is no uncomputed partition ($n_p==0$), all computed partitions are merged ($n_{um}==0$), and there is only one working thread ($n_{it}==n_t-1$), that means all partitions are computed and merged, the algorithm finishes. Otherwise, we wait until at least one thread finishes and set the state of computed partition which now only contains skylines in that partition as unmerged in Lines 13-14. In Line 15, if there are some computed and unmerged partitions, we merge each two into one new partition and set the state as uncomputed in Lines 16-17.

\begin{algorithm}[h]\small  \caption{Parallel implementation via data partitioning.} \label{Alg:dataPartition}
\SetKwInOut{Input}{input}\SetKwInOut{Output}{output}

\Input{A dataset $P$ of $n$ points in $m$ dimensions.}
\Output{Skyline of $P$.}

divide $n$ points into $s$ partitions and compute the skyline points in each partition\;
set the state of all partitions as uncomputed\;
$n_p \gets$ number of uncomputed partitions\;
$n_t \gets$ number of threads\;
$n_{it} \gets$ number of idle threads\;
$n_{um} \gets$ number of computed and unmerged results\;

\While{$n_p > 0$ \text{\normalfont \textbf{or}} $n_{it} > 0$}{
	\uIf{$n_p > 0$ \text{\normalfont \textbf{and}} $n_{it} > 0$}{
		assign one uncomputed partition to each idle thread\;
	}
        \uElse{
        		\If{$n_p == 0$  \text{\normalfont \textbf{and}}   $n_{it} == n_t- 1$  \text{\normalfont \textbf{and}}  $n_{um} == 0$ }{
			break\;
		}			
        		wait until at least one thread finishes\;
		set the state of computed partition as unmerged\;
		\If{$n_{um} > 1$}{
			merge each two into one new partition\;
			set new partition state as uncomputed\;
		}
	}
}
\end{algorithm}

\subsubsection{Discovery of Optimal Number of Partitions}\label{sec:partitionTheory}

In this subsection, we show how to calculate the optimal number of partitions for minimizing the total computation load given an independent and identically distributed random dataset. We first show the theorem of the expected number of skyline points as follows.

\begin{theorem}(\textbf{Number of Skyline Points}) \cite{DBLP:journals/jacm/BentleyKST78}.\label{The:averageNum}
Given an independent and identically distributed random dataset of $n$ points in $m$ dimensional space, the expected number of skyline points is $O(\ln^{m-1}{n})$.
\end{theorem}

In the computational complexity analysis of fully secure skyline protocol, the time complexity is $O(kn(l + m +\lceil \log n\rceil))$. According to Theorem \ref{The:averageNum}, the expected output size of input data with size $\frac{n}{s}$ in $m$ dimensional space is $\ln^{m-1}(\frac{n}{s})$. Therefore, in this step, the computation load required for each partition is $\ln^{m-1}(\frac{n}{s})\times \frac{n}{s}\times (log(\frac{n}{s})+ m + l)$. Since we have $s$ partitions, the total computation load required is $s\times ln^{m-1}(\frac{n}{s})\times \frac{n}{s}\times (\log(\frac{n}{s})+ m + l)=n\times \ln^{m-1}(\frac{n}{s})\times (\log(\frac{n}{s})+ m + l)$. This is the initial layer of the computation, which we refer to $layer_0$. We use 0 because the following layers have a slightly different formula.

\begin{figure}[htb]
 \centering
 \includegraphics[width=0.49\textwidth]{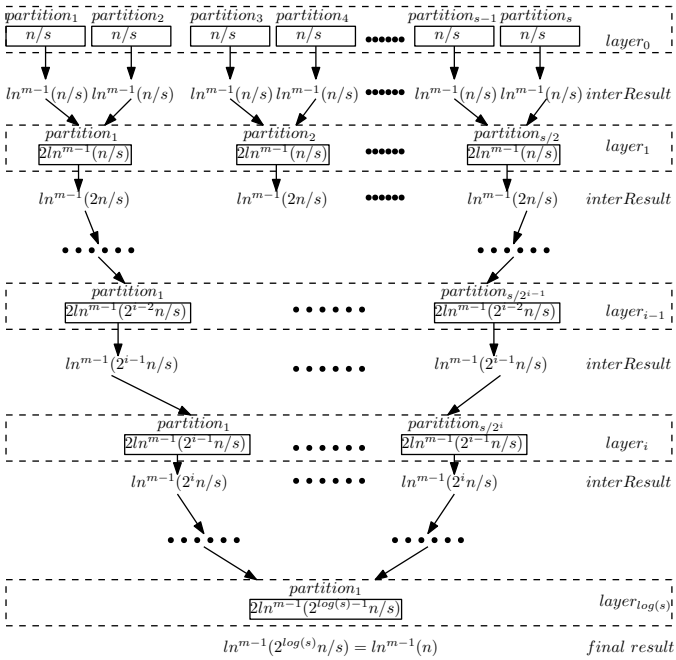}
 \caption{Layer structure (interResult is short for intermediate result).}
 \label{fig:layer}
\end{figure}

Before we proceed, we denote the number of layers excluding $layer_0$ as $n_{layer}$. For each layer $i$, we denote the number of partitions that needs to be computed as $n_{p, i}$, the size of a single input partition as $size_{in, i}$, the output size of a single partition as $size_{out, i}$, and the amount of computation load as $W_{layer_i}$. A visual graph about the layer structure is shown in Figure \ref{fig:layer}. In the ideal case, we have $s=2^h$ partitions, where $h$ is an integer. For each layer, we reduce the number of partitions by merging two partitions to form a new partition which contains skyline points of those two merged partitions. After $h$ layers' merging, we obtain only one partition which is the final skyline result.

\partitle{Number of Partitions and Layers}
To simplify the analysis, we assume the merging of two partitions happens at the same layer (although mergings from different layers may happen at the same time). As shown in Figure \ref{fig:layer}, the datasets for $layer_i$ ($i>1$) comes from the merging of two computed partitions from $layer_{i-1}$. Therefore, in $layer_i$, the number of partitions ($n_{p, i}$) is $\frac{s}{2^i}$ given the number of partitions in $layer_1$ is $\frac{s}{2}$. Meanwhile, $layer_0$ has $s$ partitions, $layer_1$ has $\frac{s}{2}$ partitions, and the last layer has one partition, so the number of layers excluding $layer_0$ ($n_{layer}$) is $\log {s}$.

\partitle{Output Size} A partition in $layer_i$ is merged from $2^i$ partitions in $layer_0$. Therefore, the expected output size of one partition at $layer_i$ corresponds to the expected output size of $2^i$ partitions in $layer_0$. That is, in $layer_i$, the expected output size of a single partition ($size_{out, i}$) is $ln^{m-1}(\frac{2^{i}n}{s})$.

\partitle{Input Size} In $layer_i$, the size of each input partition ($size_{in, i}$) is twice of the single partition output size from the last layer because it is the merging of two outputs from the last layer. In other words, $size_{in, i} = 2 \times size_{out, i-1}$ = $2 \times ln^{m-1}(\frac{2^{i-1}n}{s})$. For example, the expected single partition output size of $layer_0$ is $\ln^{m-1}(\frac{n}{s})$ and the expected size of each input partition in $layer_1$ is $2\times \ln^{m-1}(\frac{n}{s})$.

\partitle{Computation Load} With $n_{p, i}$, $size_{in, i}$, and $size_{out, i}$, we can obtain the general formula for computation load of $layer_i$ ($i \neq $ 0) as $W_{layer_i}$ = $n_{p, i} \times size_{out, i} \times size_{in, i} \times (m + \log (size_{in, i}))$ according to the time complexity of our fully secure skyline protocol. And since we have $n_{layer}$ layers, the overall computation load is calculated as follows.

\begin{displaymath}
\begin{split}
W_{all}=&W_{layer_0} + \sum_{1}^{n_{layer}}W_{layer_i} \\
=&W_{layer_0} + \sum_{1}^{n_{layer}} n_{p, i} \times size_{out, i} \times size_{in, i} \times (m + \log (size_{in, i})) \\
=&n\times \ln^{m-1}(\frac{n}{s})\times (\log{\frac{n}{s}}+ m + l )+
\sum_{i=1}^{\log{s}}\frac{s}{2^{i}} \times \\
&\ln^{m-1}(\frac{2^in}{s})\times
2 ln^{m-1}(\frac{2^{i-1}n}{s})
\times (\log(2 ln^{m-1}(\frac{2^{i-1}n}{s}))+m + l)
\end{split}
\end{displaymath}

\partitle{Optimal Number of Partitions}
Without loss of generality, from now on, we assume $n=2^u$ and $s=2^v$, where $u,v\in \mathbb{Z^{+}}$ and $1\leq v < u$. To find out the optimal number of partitions, our goal is to minimize $W_{all}$ against $s$ or $v$. Because $n=2^u$ and $s=2^v$, we have the computation load $W(v)$ corresponding to the number of partition $s=2^v$ as follows.

\begin{displaymath}
\begin{split}
W(v) = & 2^u \times (u-v)^{m-1} \times \ln^{m-1}2 \times (u-v+ m+l)+ \\
& \sum_{i=1}^{v} 2^{v-i+1} \times
(i+u-v)^{m-1}
\times (i-1+u-v)^{m-1} \times \ln^{2m-2}2 \\
& \times (\log(2\times (i-1+u-v)^{m-1}\ln^{m-1}2)+m+l)
\end{split}
\end{displaymath}

We denote the part after $\sum$ as $WI_{v, i}$. Notice that $WI_{v, i}=WI_{v+1, i+1}$, we have
\begin{displaymath}
\begin{split}
W(v+1)-W(v) = & W_{layer_0, v+1} - W_{layer_0, v} + \sum_{i=1}^{v+1} WI_{v+1, i} - \sum_{i=1}^{v} WI_{v, i} \\
= & W_{layer_0, v+1} - W_{layer_0, v} + WI_{v+1, 1} 
\end{split}
\end{displaymath}

Notice that the minimal value of $W$ lies at the position where $W(v+1)-W(v)$ changes from negative to positive. Observe that in our setting, all variables can only be positive integer, which means we need to find out the integer $v$ such that $f(v) = W(v+1)-W(v)$ changes from negative to positive. By letting $x=u-v$, we have
\begin{displaymath}
\begin{split}
f(x) = & WI_{v+1, 1} + W_{layer_0, v+1} - W_{layer_0, v} \\
= & 2^{v+1} \times (x)^{m-1}
\times (x-1)^{m-1} \times \ln^{2m-2}2 \\
& \times (\log(2\times (x-1)^{m-1}\ln^{m-1}2)+m+l) \\
& + 2^u \times (x-1)^{m-1} \times \ln^{m-1}2 \times (x-1+m+l) \\
& - 2^u \times x^{m-1} \times \ln^{m-1}2 \times (x+m+l) \\
= & 2^{u} \ln^{m-1}2 \times (
2^{1-x} \times x^{m-1} \times (x-1)^{m-1} \times \ln^{m-1}2 \\
& \times (\log(2\times (x-1)^{m-1}\ln^{m-1}2)+m+l) \\
& + ( (x-1)^{m-1} \times (x-1+m+l)
- x^{m-1} \times (x+m+l)) )
\end{split}
\end{displaymath}

To obtain the minimal value of $f(x)$, we can ignore the preceding $2^{u}\ln^{m-1}2$ which is always positive. Then we can easily solve the problem to find out $x$ where $f(x)$ changes from positive to negative given $m$ and $l$.

For example, we set $l=20$ in our experiments, if $m=2$, then the minimal value of $W(v)$ is obtained at $x=1$, i.e., $u-v=1$. This actually corresponds to the case where each initial partition has two data points. If $m=3$, we have $x=6$, i.e., $u-v=6$. That is, for three dimensional datasets, the optimal number of partitions is $2^{u-6}$ and each partition has $2^6$ points.

\subsection{Optimization of Lazy Merging}\label{subsec:lazymerge}

In this subsection, we show another optimization with lazy merging.

\partitle{Lazy Merging} In the hierarchical divide-and-conquer approach proposed in the last subsection, results from any two computed partitions are merged immediately as a new partition for computing skyline points. However, immediate merging might not be optimal in the later stage of the program because it requires 1) more merging overhead and 2) more unnecessary computations. In the later stage of the program, there are only a few points in each partition. At this time, merging overhead is high compared to the computation time. Therefore, we can employ lazy merging which incurs less merging overhead. Furthermore, in the later stage of the program, those remaining points are likely to follow an anti-correlated distribution as they are skyline points of a partition at a previous layer. For anti-correlated dataset, data partitioning will incur more unnecessary computations. Consider an extreme example, if all the remaining points are the final skyline points, all the computations in each partition are unnecessary. Therefore, we can employ lazy merging to avoid those unnecessary computations and delay the merging operation to a later time when more computed results are ready.

\partitle{Merging Timing} With lazy merging, we can reduce running time if and only if the timing for lazy merging is perfect. Merging too early (immediate merging) or merging too late does not provide enough benefit or even jeopardizes the performance. As shown in the last subsection, for a given dataset, we can calculate the optimal number of partitions, which is related to the dataset size. For example, given $l=20$ and $m=3$, we have the number of optimal partitions as $\frac{n}{2^6}$, which effectively states that the optimal size of each partition should be $2^6=64$ in the initial layer. Therefore, in our algorithm, we heuristically wait until the size of merged partitions reach $64$ before sending it for computation in the previous example. That is, there are at least $64$ points in each partition (excluding the final partition which contains the final skyline points) to compute the skyline points.

\partitle{Security Analysis} The cloud servers can tell if the subsets are skew or uniformly distributed in the extreme case when the distribution of entire dataset is different from the distribution of subsets based on the different number of returned skyline points from each partition. However, the probability is very low because we randomly partition the dataset, and the distribution of subsets should be very similar to the distribution of entire dataset. Moreover, this attack can be easily fixed by returning all the tuples in each iteration. That is, cloud servers $\mathcal{C}_1$ and $\mathcal{C}_2$ return all skyline tuples with true values and non-skyline tuples with MAX values. In this way, the cloud servers cannot know the skyline distribution of subsets, thus, the cloud servers cannot get any new information from the partitions.

\section{Experiments}\label{sec:experiments}

In this section, we describe our experimental setup and optimized parallel system design. For comparison purposes, we have implemented both protocols: the Basic Secure Skyline Protocol (\textbf{BSSP}) in Section \ref{subsec:basicProtocol}, and the Fully Secure Skyline Protocol (\textbf{FSSP}) in Section \ref{subsec:fullyProtocol}. Since there is no existing solution for secure skyline computation, we use the basic approach as a baseline which is efficient but leaks some indirect data patterns to the cloud server. We have also designed a parallel framework for effective reducing computation time together with the two optimizations, data partitioning and lazy merging.

\subsection{Experiment Setup}
We implemented all algorithms in C with all multithreading using POSIX threads and all communication using sockets. We ran single-machine-experiments on a machine with Intel Core i7-6700K 4.0GHz running Ubuntu 16.04. The distributed version was tested on a cluster of 64 machines with Intel Core i7-2600 3.40GHz running CentOS 6, which we will provide more details in the next section.
In our experiment setup, both $\mathcal{C}_1$ and $\mathcal{C}_2$ were running on the same machine. The reported computation time is the total computation time of the $\mathcal{C}_1$ and $\mathcal{C}_2$.

\partitle{Datasets}
We used both synthetic datasets and a real NBA dataset in our experiments. To study the scalability of our methods, we generated independent (INDE), correlated (CORR), and anti-correlated (ANTI) datasets following the seminal work \cite{DBLP:conf/icde/BorzsonyiKS01}. We also built a dataset that contains 2384 NBA players who are league leaders of playoffs\footnote{The data was extracted from http://stats.nba.com/leaders /alltime/?ls=iref:nba:gnav on 04/15/2015}. Each player has five attributes that measure the player's performance: Points (PTS), Rebounds (REB), Assists (AST), Steals (STL), and Blocks (BLK).

\partitle{Data Partitioning}
This procedure can be done either using single thread or multiple threads. We conducted single thread experiment for verifying the optimal number of partitions. And we refer to multithreading implementation as local parallelism. The algorithm is shown in Algorithm \ref{Alg:dataPartition}.

To further demonstrate the scalability of our algorithm, we also implemented a distributed version, which employs a manager-worker model. The manager distributes partitions to workers, the workers compute the skyline points in any given dataset and return the results to the manager, which works similarly as the local parallelism. The only difference is that the manager could implement sophisticated load balancing algorithm to fully utilize the computation resources. The overall data partitioning scheme is very similar to the existing MapReduce approach. However, we didn't employ existing MapReduce framework because existing crypto library in Java does not satisfy our requirements.

\partitle{Lazy Merging}
The lazy merging delays the merging operation until there are enough results to form a partition with optimal size, which is detailed shown in Section~\ref{sec:partitionTheory}. All experiments using optimizations are conducted using $10$ different independent and identically distributed random datasets of size $512$ and dimension $3$ with three repeated runs for each dataset.

\subsection{Impact of Parameters}\label{sec:eval-params}

In this subsection, we evaluate our protocols by varying the number of tuples (n), the number of dimensions (m), and the key size (K) on datasets of various distributions.

\begin{figure*}[!htb]
\centering
\subfigure[time cost of CORR]{
\begin{minipage}[b]{0.23\textwidth}
\includegraphics[width=1.13\textwidth]{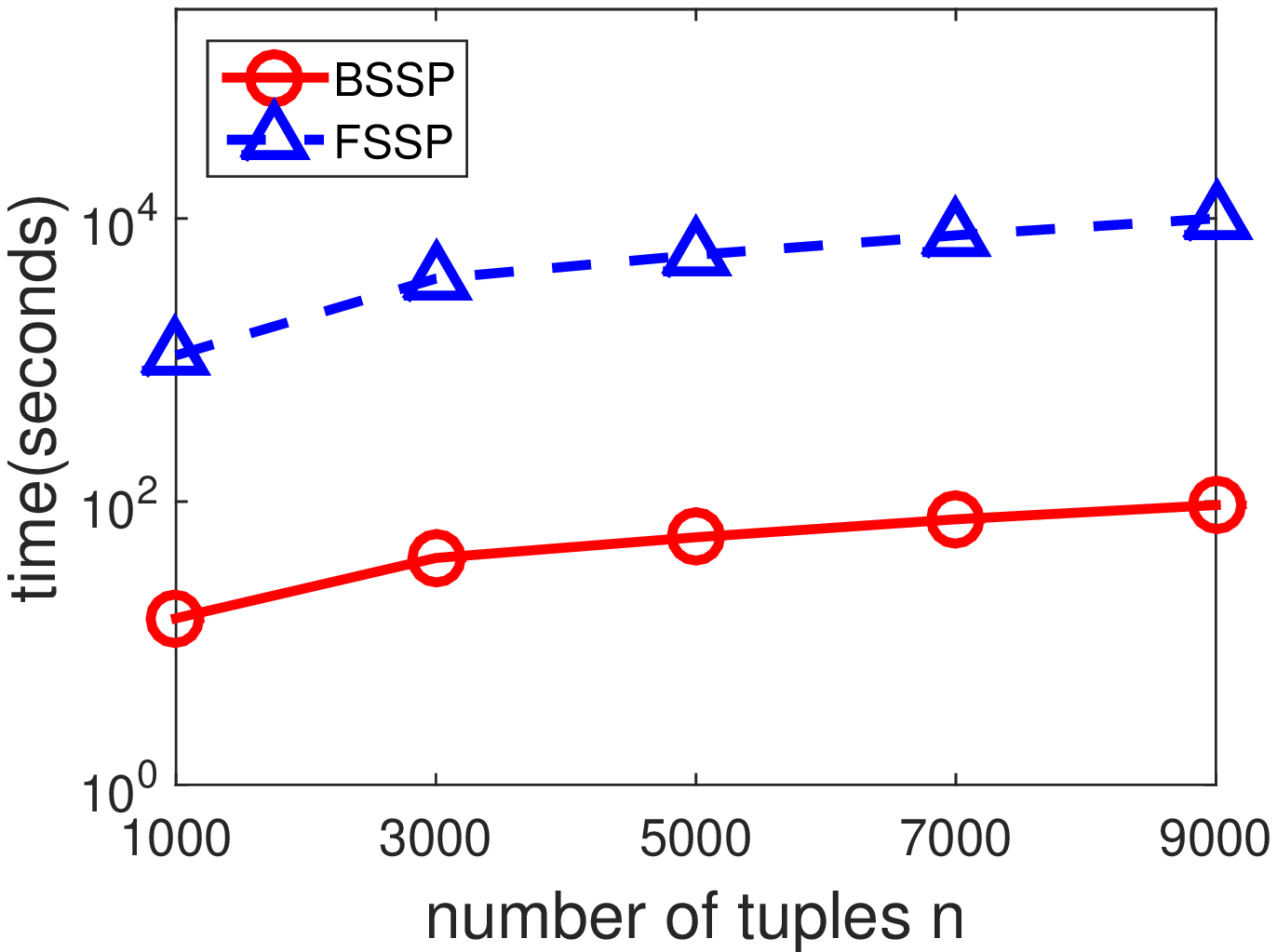}
\end{minipage}
}
\subfigure[time cost of INDE]{
\begin{minipage}[b]{0.23\textwidth}
\includegraphics[width=1.13\textwidth]{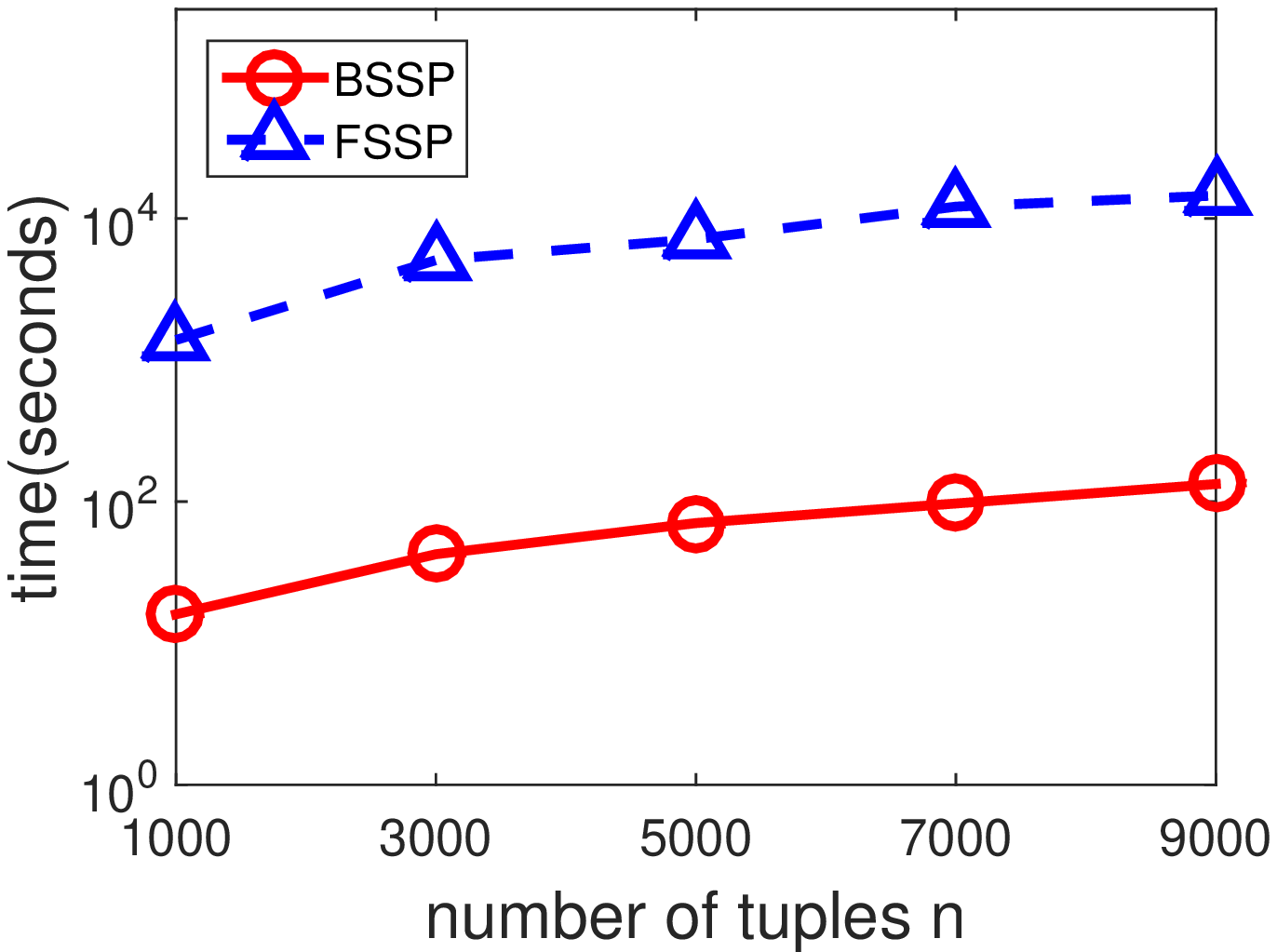}
\end{minipage}
}
\subfigure[time cost of ANTI]{
\begin{minipage}[b]{0.23\textwidth}
\includegraphics[width=1.13\textwidth]{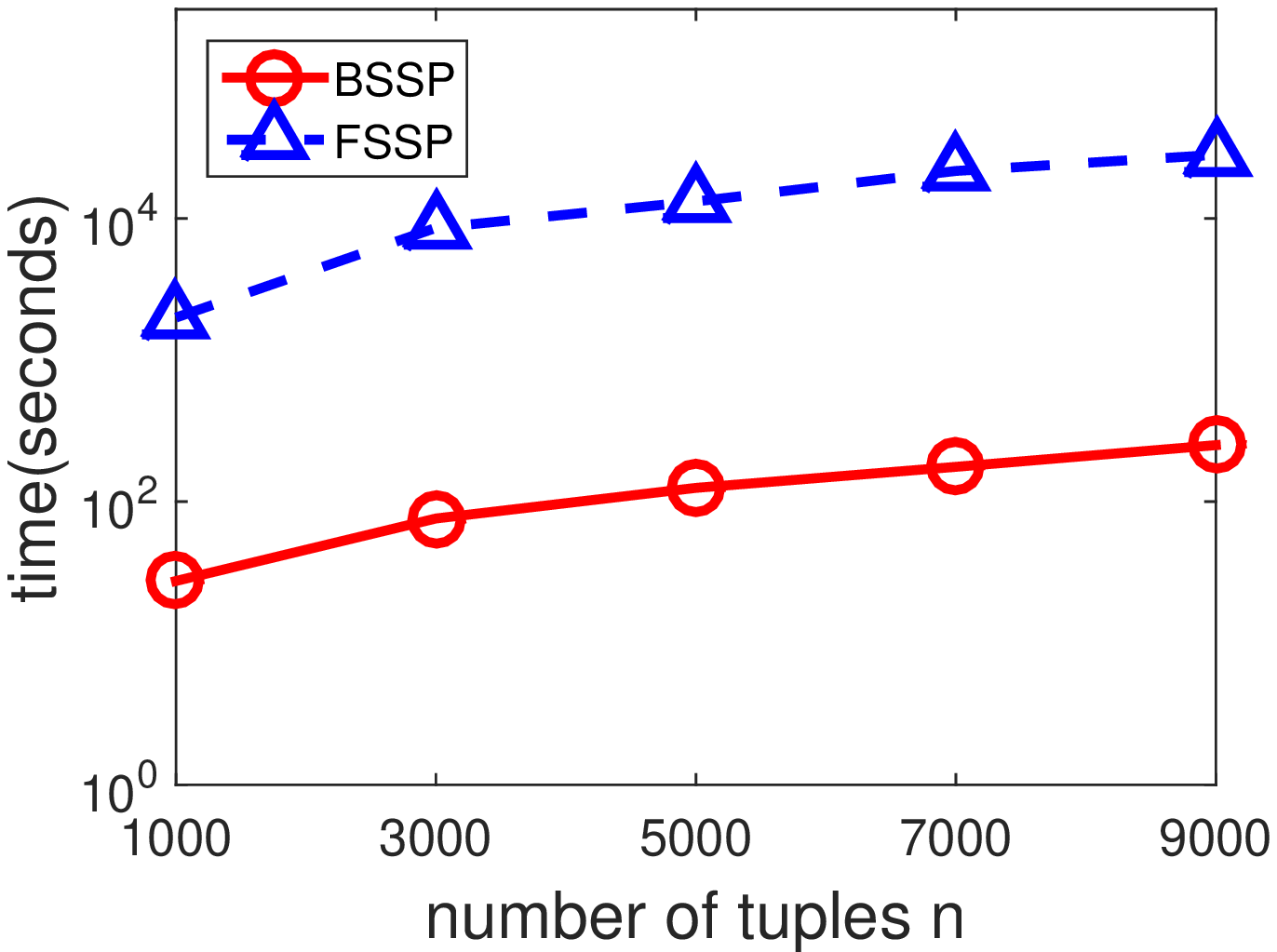}
\end{minipage}
}
\subfigure[time cost of NBA]{
\begin{minipage}[b]{0.23\textwidth}
\includegraphics[width=1.13\textwidth]{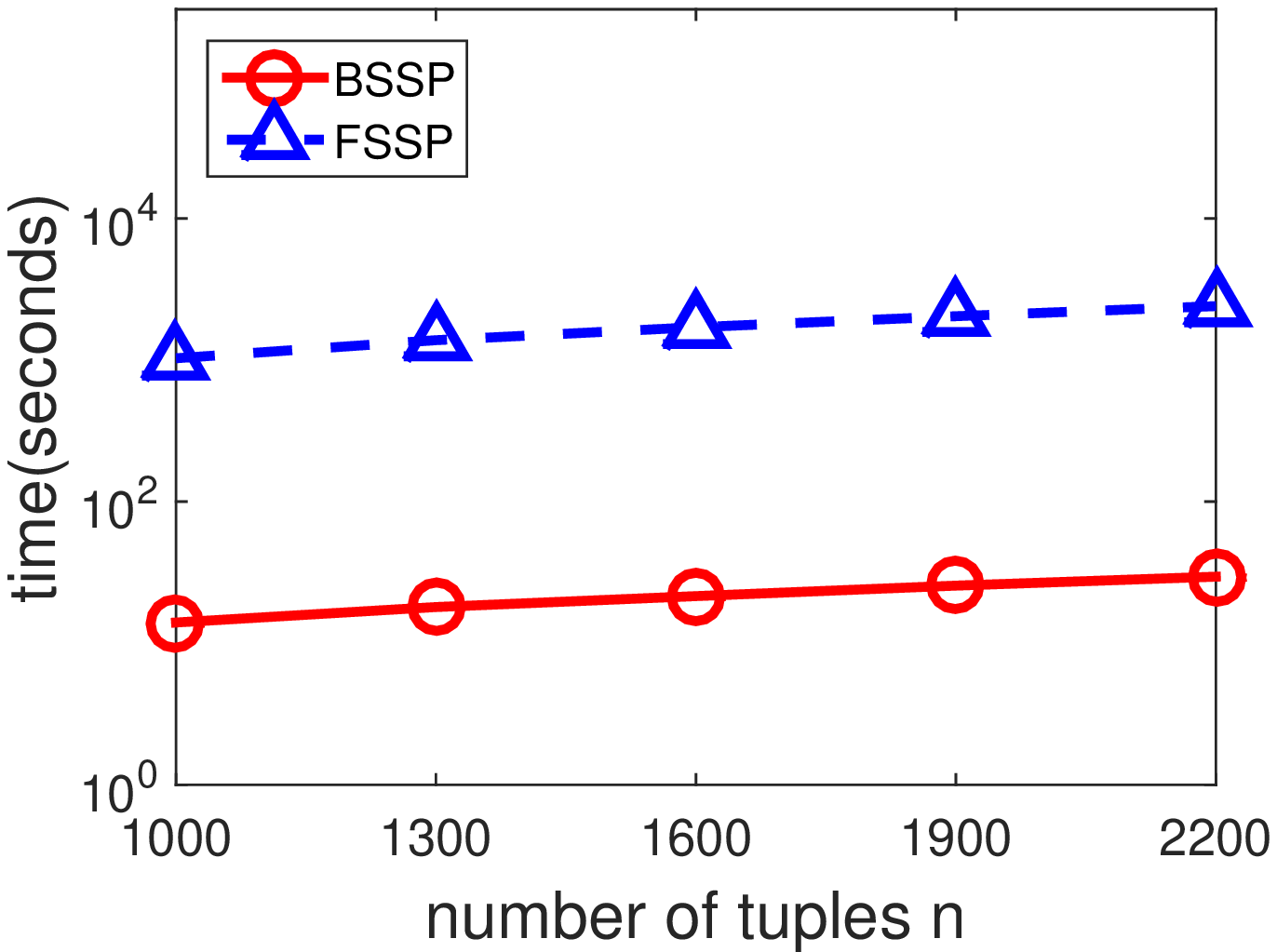}
\end{minipage}
}
\vspace{-1em}
\caption{The impact of n (m=2, K=512).} \label{fig:diffn}
\end{figure*}

\partitle{Impact of number of tuples $n$} Figures \ref{fig:diffn}(a)(b)(c)(d) show the time cost of different $n$ on CORR, INDE, ANTI, and NBA datasets, respectively. We observe that for all datasets, the time cost increases approximately linearly with the number of tuples $n$, which is consistent with our complexity analysis.  While BSSP is very efficient, FSSP does incur more computational overhead for full security.  Comparing different datasets, the time cost is in slightly increasing order for CORR, INDE, and ANTI, due to the increasing number of skyline points of the datasets. The time for NBA dataset is low due to its small number of tuples.

\begin{figure*}[!htb]
\centering
\subfigure[time cost of CORR]{
\begin{minipage}[b]{0.23\textwidth}
\includegraphics[width=1.13\textwidth]{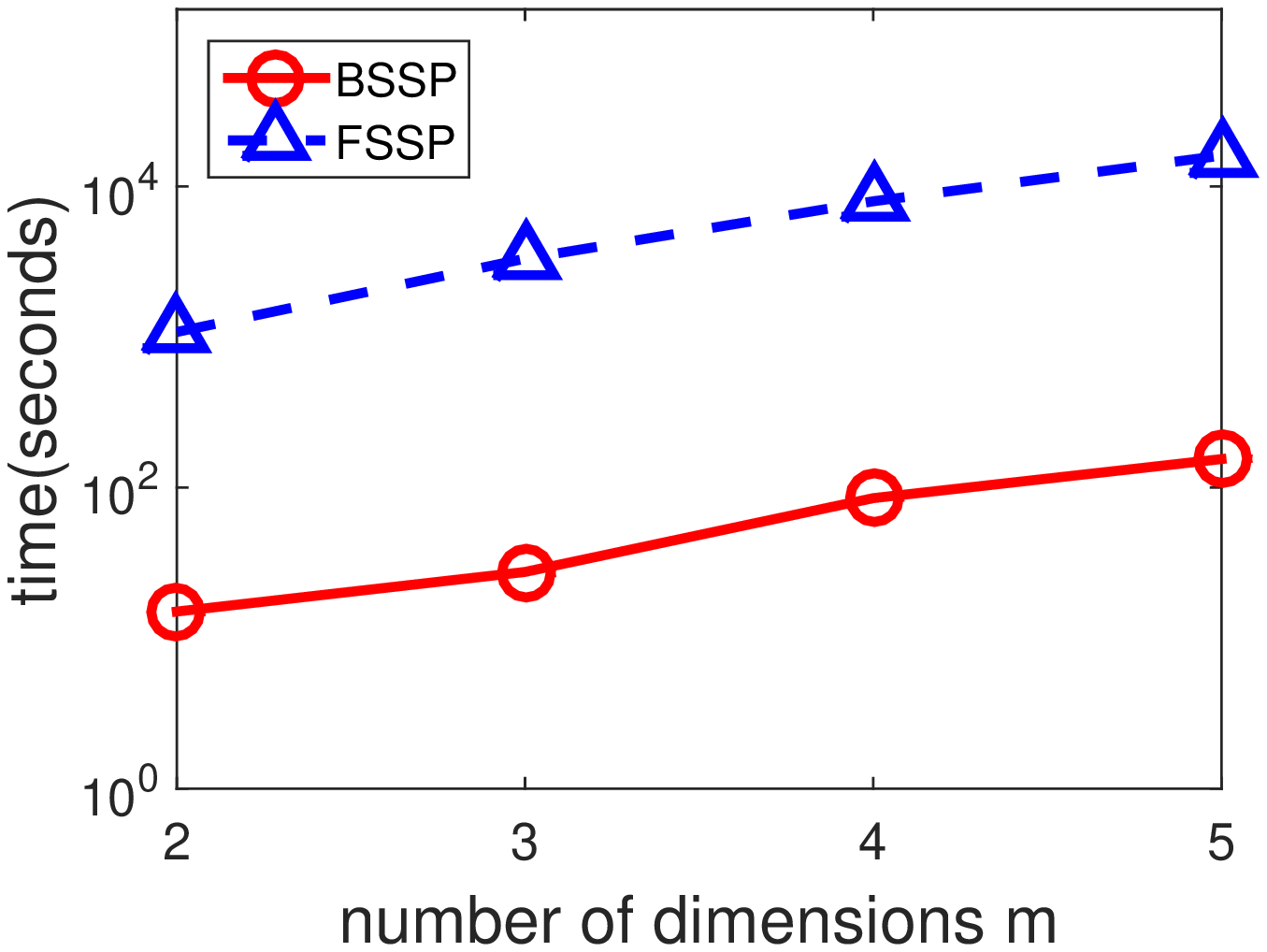}
\end{minipage}
}
\subfigure[time cost of INDE]{
\begin{minipage}[b]{0.23\textwidth}
\includegraphics[width=1.13\textwidth]{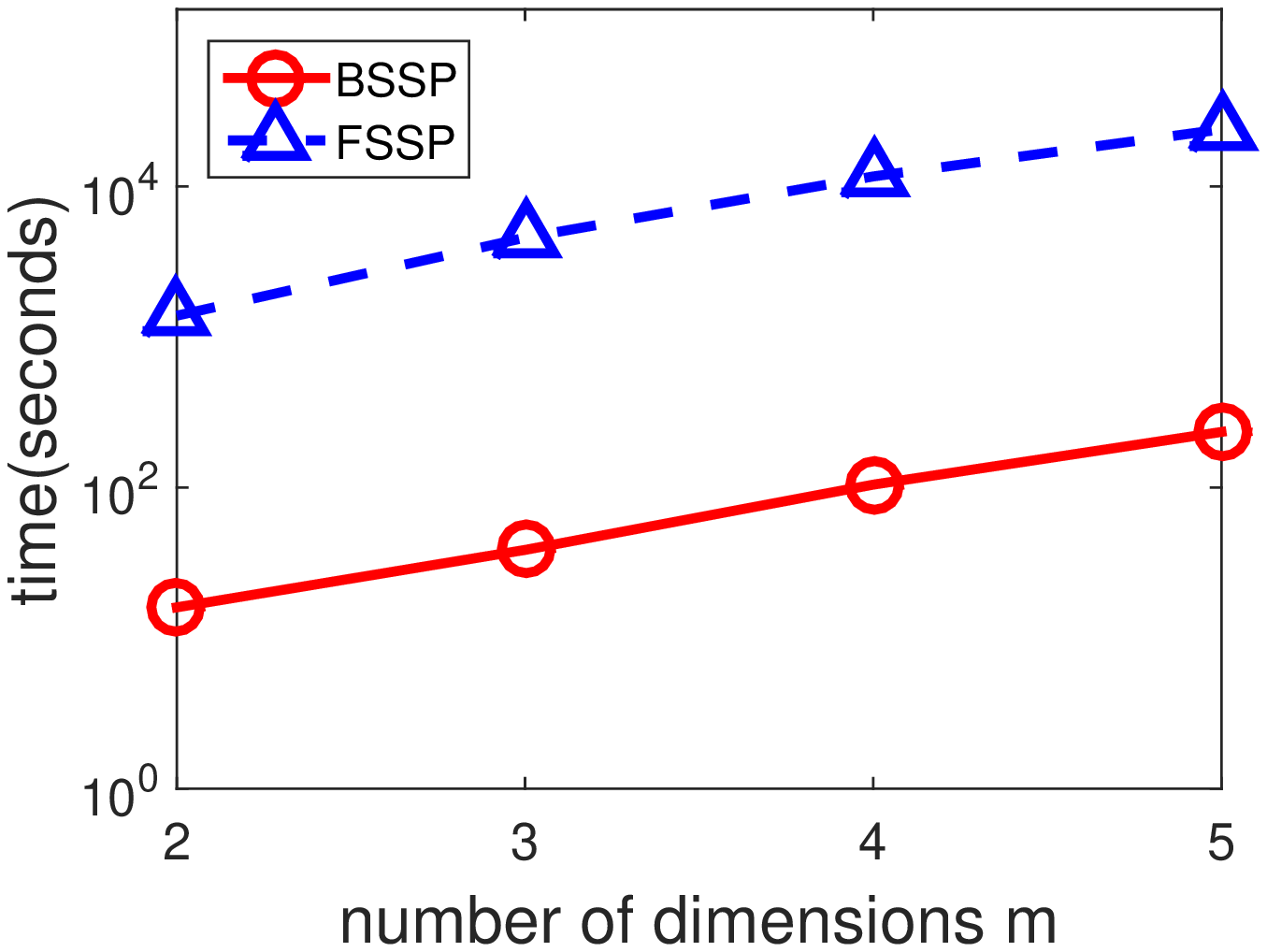}
\end{minipage}
}
\subfigure[time cost of ANTI]{
\begin{minipage}[b]{0.23\textwidth}
\includegraphics[width=1.13\textwidth]{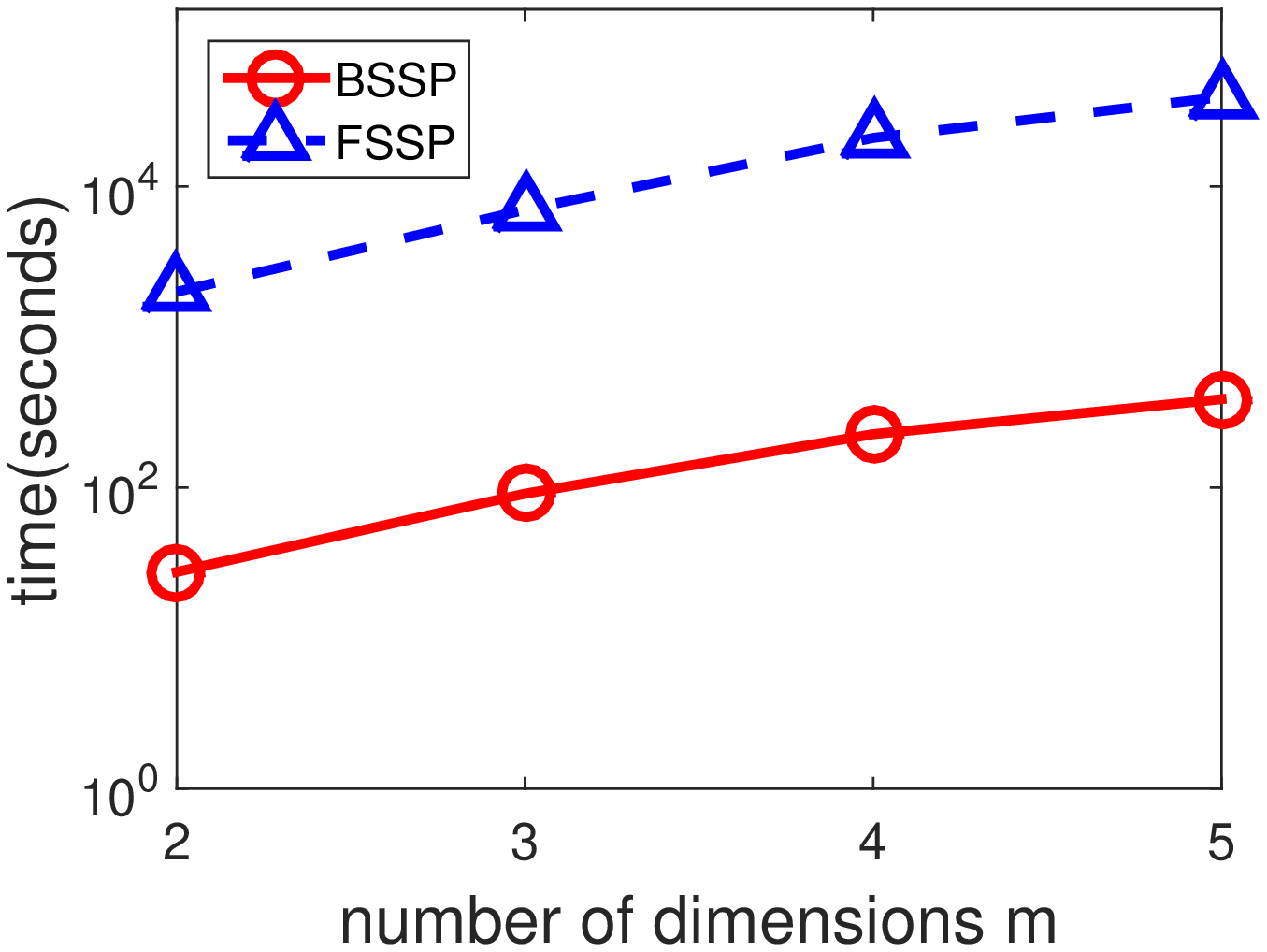}
\end{minipage}
}
\subfigure[time cost of NBA]{
\begin{minipage}[b]{0.23\textwidth}
\includegraphics[width=1.13\textwidth]{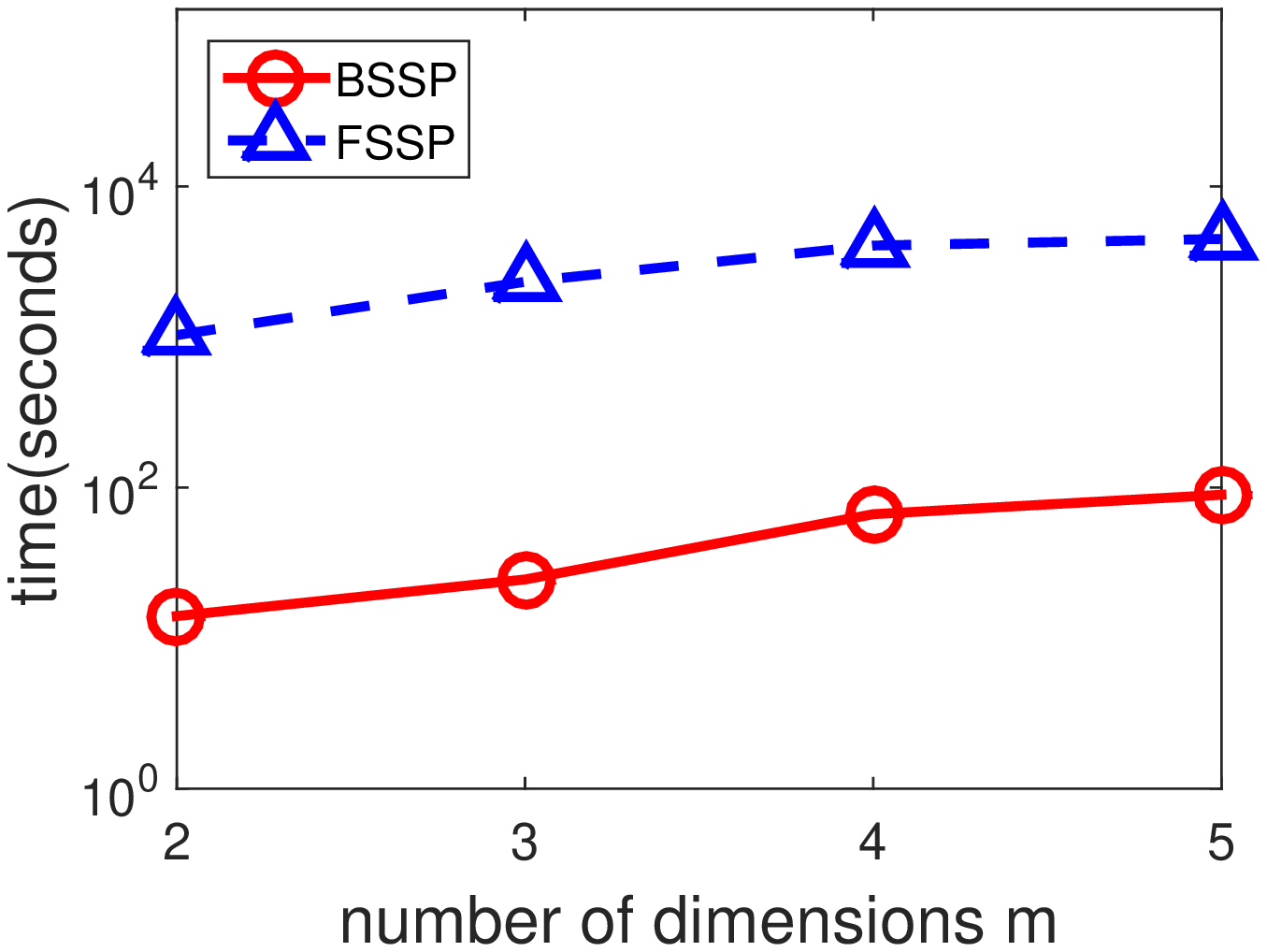}
\end{minipage}
}
\vspace{-1em}
\caption{The impact of m (n=1000, K=512).} \label{fig:diffm}
\end{figure*}

\partitle{Impact of number of dimensions $m$} Figures \ref{fig:diffm}(a)(b)(c)(d) show the time cost of different $m$ on CORR, INDE, ANTI, and NBA datasets, respectively. For all datasets, the time cost increases  approximately linearly with the number of dimensions $m$. FSSP also shows more computational overhead than BSSP.  The different datasets show a similar comparison as in Figure \ref{fig:diffn}.  The time for NBA dataset is lower than the CORR dataset which suggests that the NBA data is strongly correlated.

\begin{figure*}[!htb]
\centering
\subfigure[time cost of CORR]{
\begin{minipage}[b]{0.23\textwidth}
\includegraphics[width=1.13\textwidth]{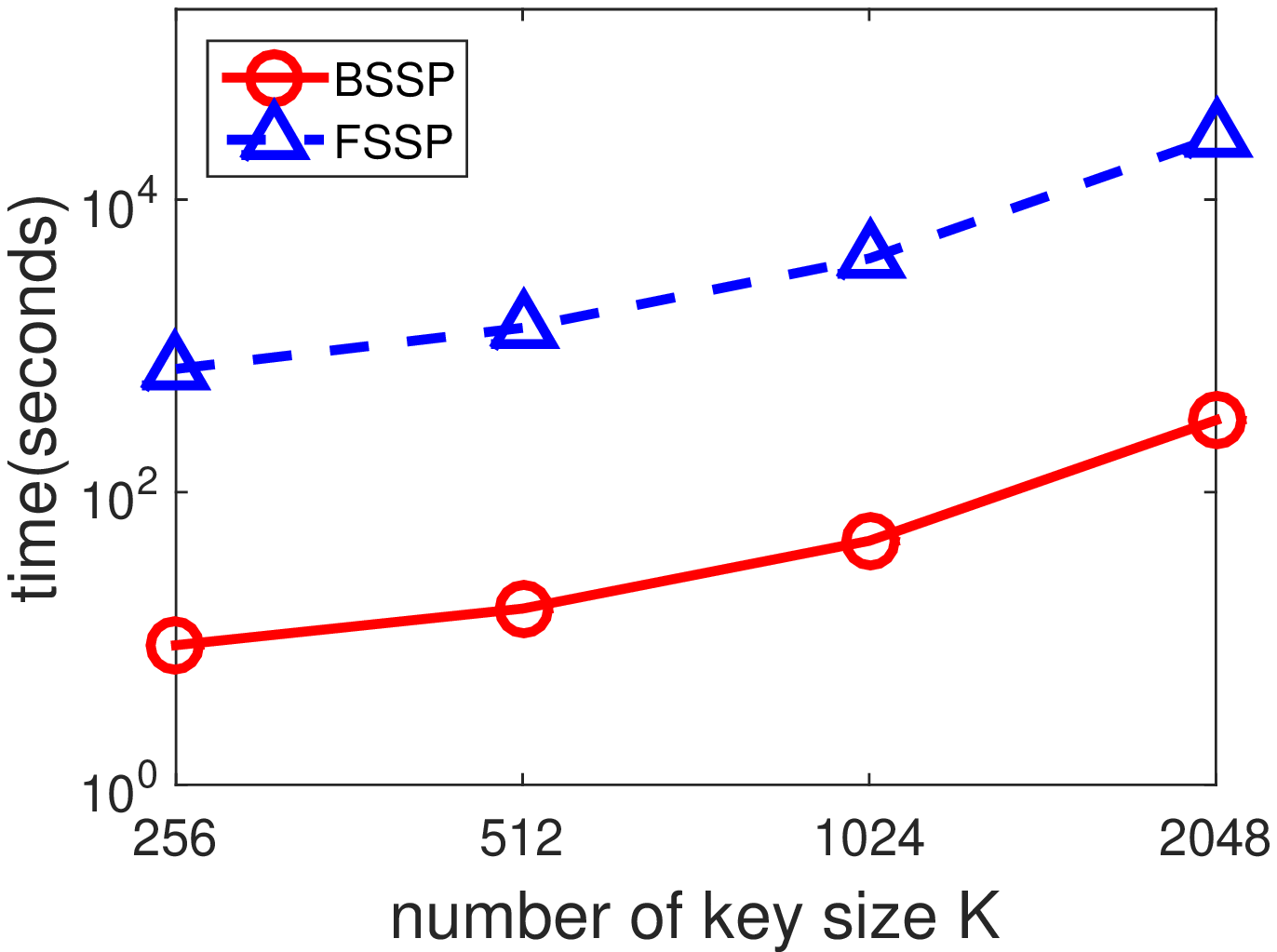}
\end{minipage}
}
\subfigure[time cost of INDE]{
\begin{minipage}[b]{0.23\textwidth}
\includegraphics[width=1.13\textwidth]{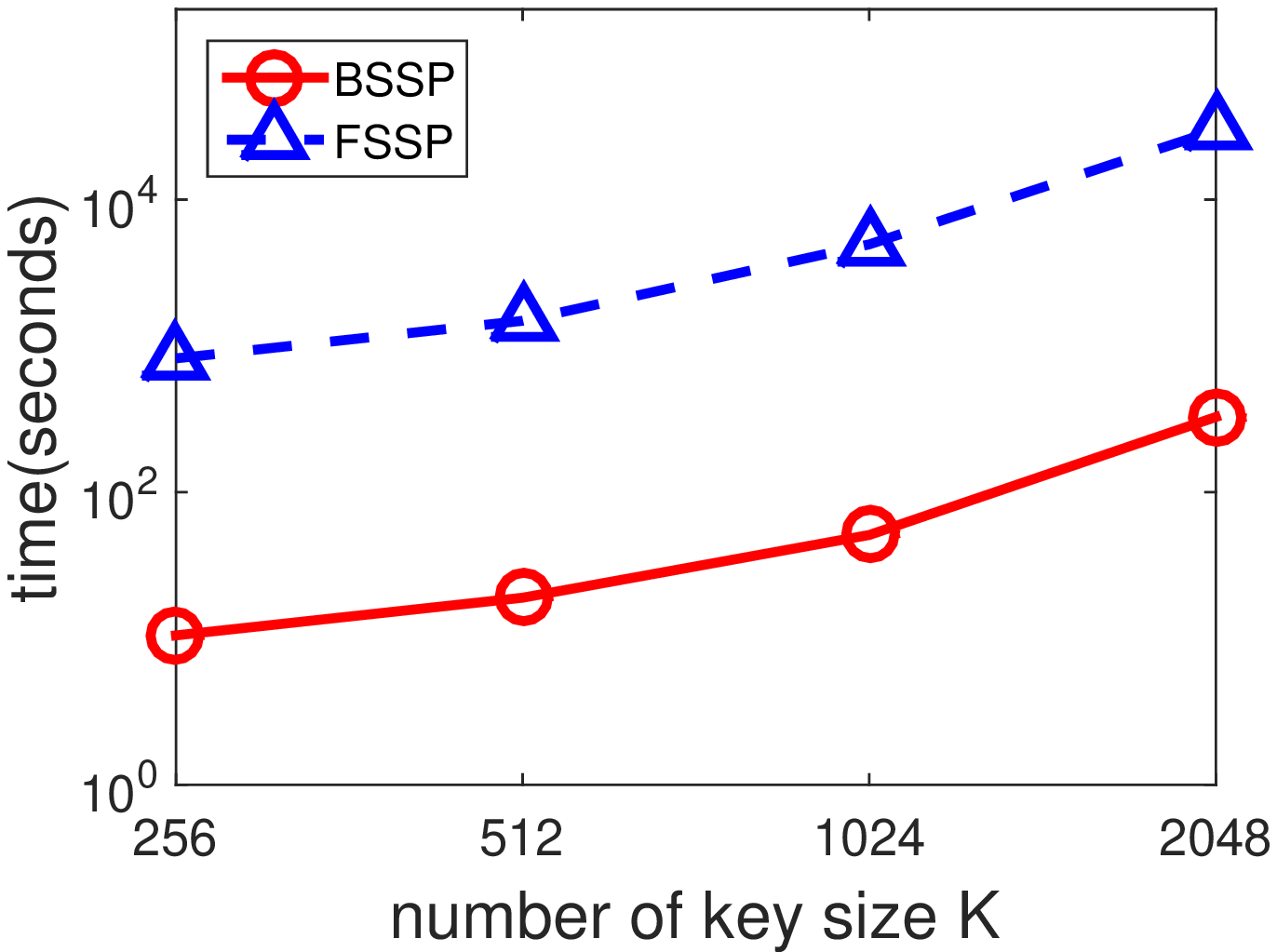}
\end{minipage}
}
\subfigure[time cost of ANTI]{
\begin{minipage}[b]{0.23\textwidth}
\includegraphics[width=1.13\textwidth]{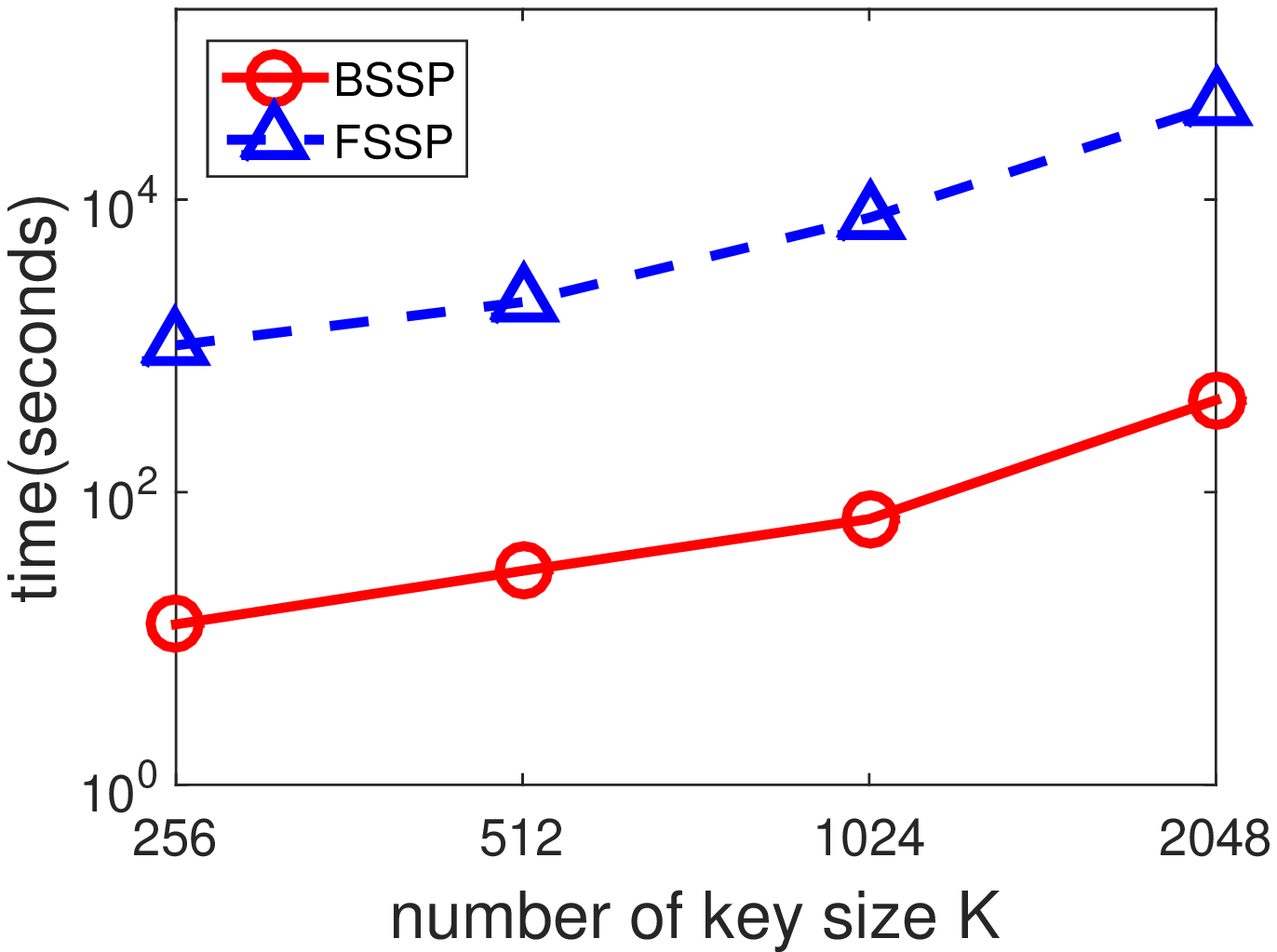}
\end{minipage}
}
\subfigure[time cost of NBA]{
\begin{minipage}[b]{0.23\textwidth}
\includegraphics[width=1.13\textwidth]{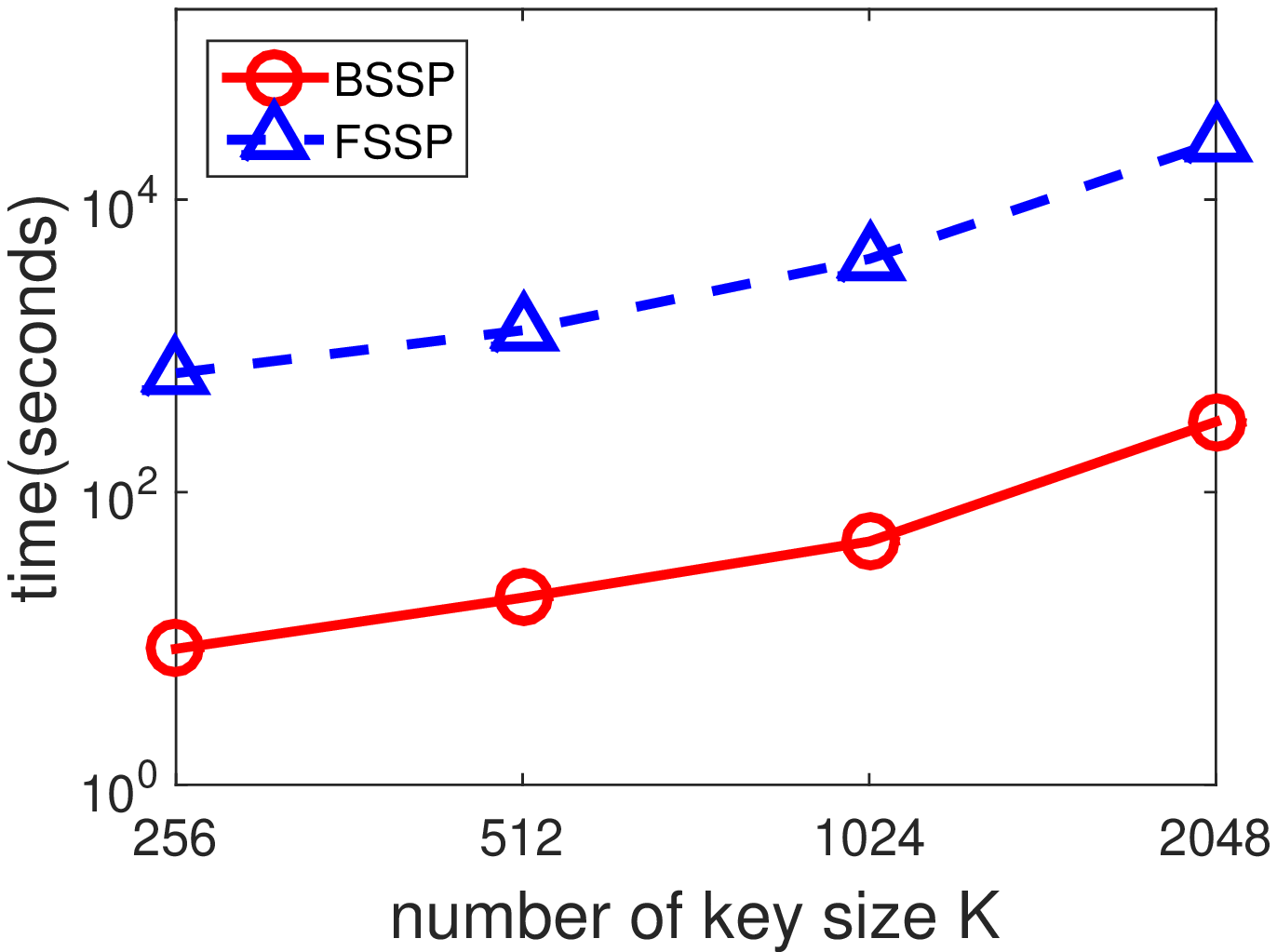}
\end{minipage}
}
\vspace{-1em}
\caption{The impact of K (n=1000, m=2).} \label{fig:diffkey}
\end{figure*}

\partitle{Impact of encryption key size $K$} Figures \ref{fig:diffkey}(a)(b)(c)(d) show the time cost with different key size used in the Paillier cryptosystem on CORR, INDE, ANTI, and NBA datasets, respectively.  A stronger security indeed comes at the price of computation overhead, i.e., the time cost increases significantly, almost exponential, when $K$ grows.

\begin{figure}[h]
 \centering
 \includegraphics[width=0.3\textwidth]{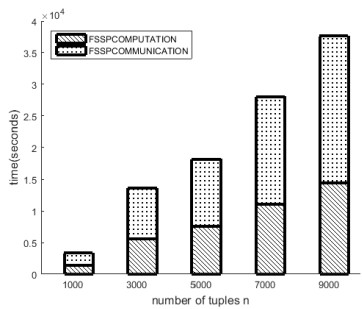}
 \caption{Computation and communication time cost of different n (m=2, K=512).}
 \label{fig:ncomm}
\end{figure}

\partitle{Communication overhead} We also measured the overall time which includes computation time reported earlier and the communication time between the two server processes. Figure \ref{fig:ncomm} shows the computation and communication time of different $n$ on INDE dataset of FSSP. We observe that computation time only takes about one third of the total time in this setting.

\subsection{Effect of Optimizations}\label{sec:eval-opt}
In this subsection, we evaluate the efficiency of our proposed two optimizations, data partitioning and lazy merging.

\begin{figure}[t]
\begin{minipage}[t]{0.22\textwidth}
\centering
\includegraphics[width=1.05\textwidth]{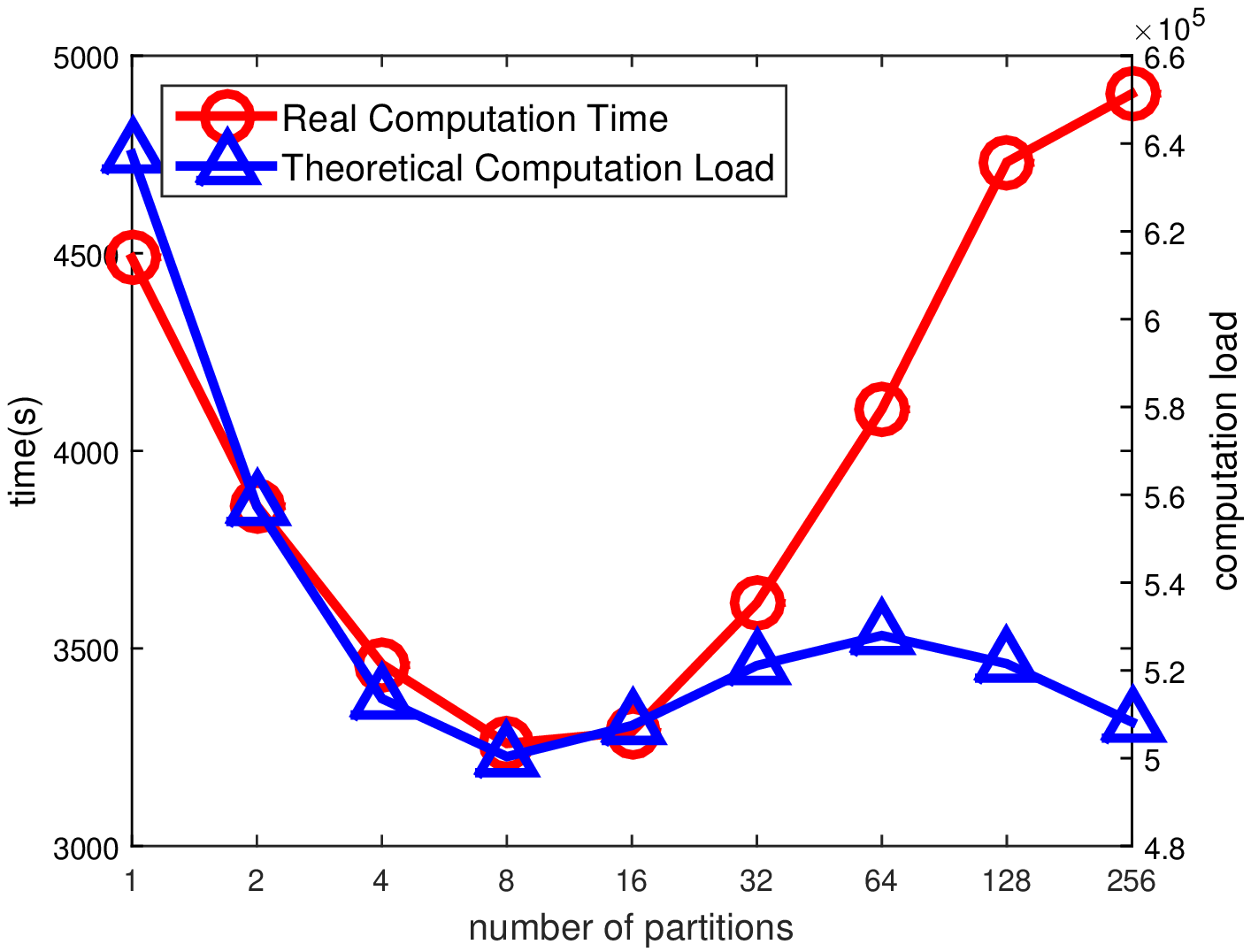}
\caption{Theoretical and experimental results.}
\label{fig:D3_avg_no}
\end{minipage}%
\begin{minipage}[t]{0.22\textwidth}
\centering
\includegraphics[width=1.05\textwidth]{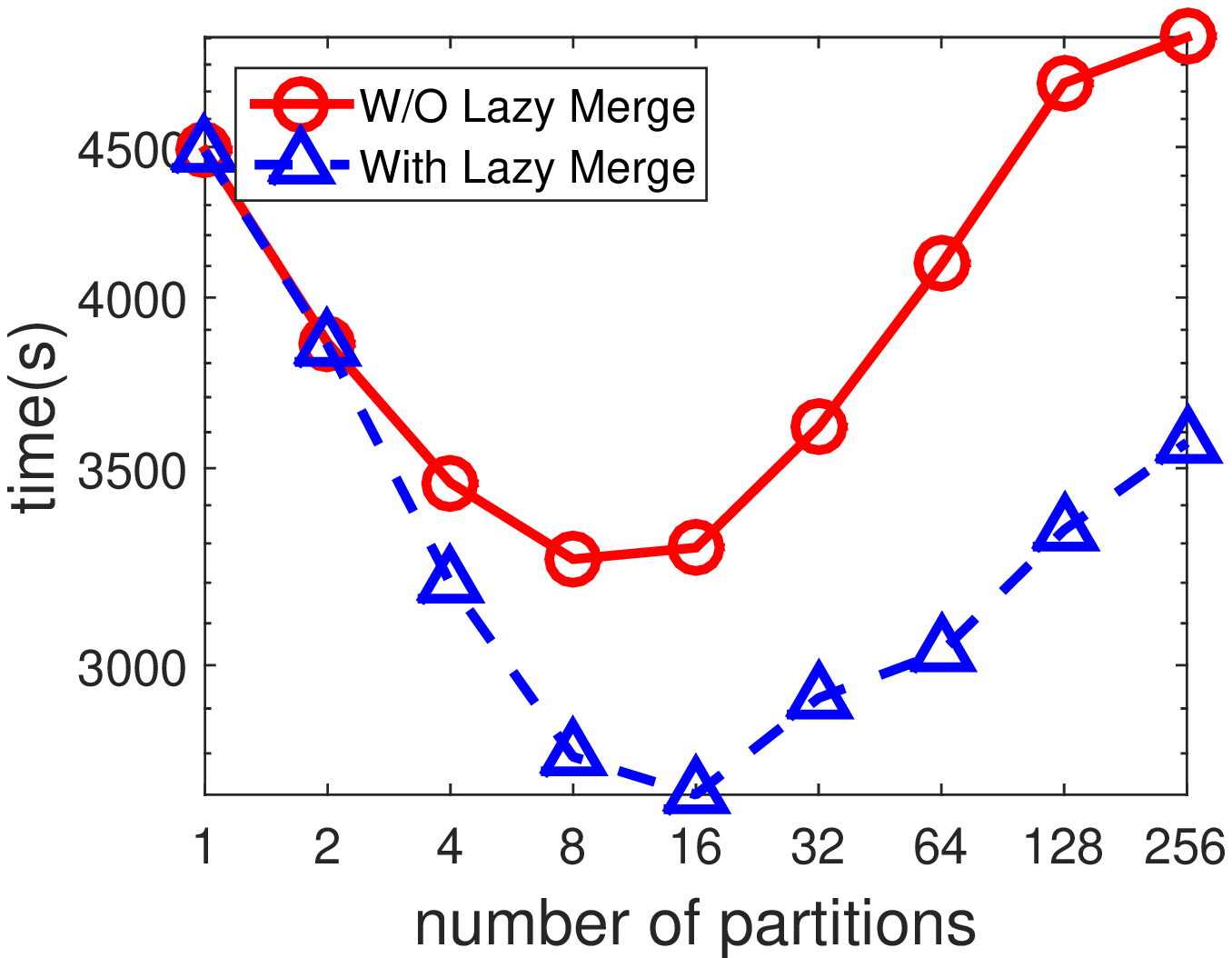}
\caption{Computation time with and without lazy merging.}
\label{fig:D3_avg_yes}
\end{minipage}
\end{figure}

\partitle{Data Partitioning} Figure \ref{fig:D3_avg_no} shows the relationship between theoretical computation load and real computation time. The theoretical computation load has an optimal value at the partition $2^{9-6}=8$, which indicates dividing the original dataset into $8$ partitions will give the smallest amount of computation load. Using ten datasets and three repeated runs for each dataset, we obtained the real computation time, which perfectly matches the theoretical computation load at the region with small number of partitions. With large number of partitions, the experimental results deviate from theoretical derivations. The reason for the deviation is that when the number of points in each partition is too small for large number of partitions, the number of skyline points in each partition violates our assumption of data distribution. For example, it is hard to say a dataset with only five points is an independent and identically distributed random dataset. Therefore, computation time for each partition does not follow our derivation. Furthermore, the large number of partitions will incur more merging overhead.

\partitle{Lazy Merging} As yet another optimization, lazy merging plays an important role especially when the number of partitions is large. In Figure \ref{fig:D3_avg_yes}, we show the computation time with and without lazy merging, respectively. It can be seen that overall with lazy merging, the run time can be effectively reduced. The larger number of partitions, the larger number of time difference, which is reasonable because the larger number of partitions, the larger number of merging operations and more rounds of computation. We can also see that for one partition (no partition) and two partitions, there is no time reduction, the reasons are that there is no merging operation need for one partition and there is no lazy merging operation for two partitions.

To summarize, both data partitioning and lazy merging have been proven effective and can significantly reduce the computation time even using single thread.

\subsection{Effect of Parallelism}\label{sec:eval-scalability}
In this subsection, we demonstrate the speedup of our protocol by using multithreading (local parallelism) on independent and identically distributed random datasets with $512$ points and distributed computing with $64$ commercial desktops (global parallelism) on independent and identically distributed random datasets with $65536$ points.

\begin{figure}[!htb]
\centering
\subfigure[Local parallelism.]{
\begin{minipage}[b]{0.22\textwidth}
\includegraphics[width=1.15\textwidth]{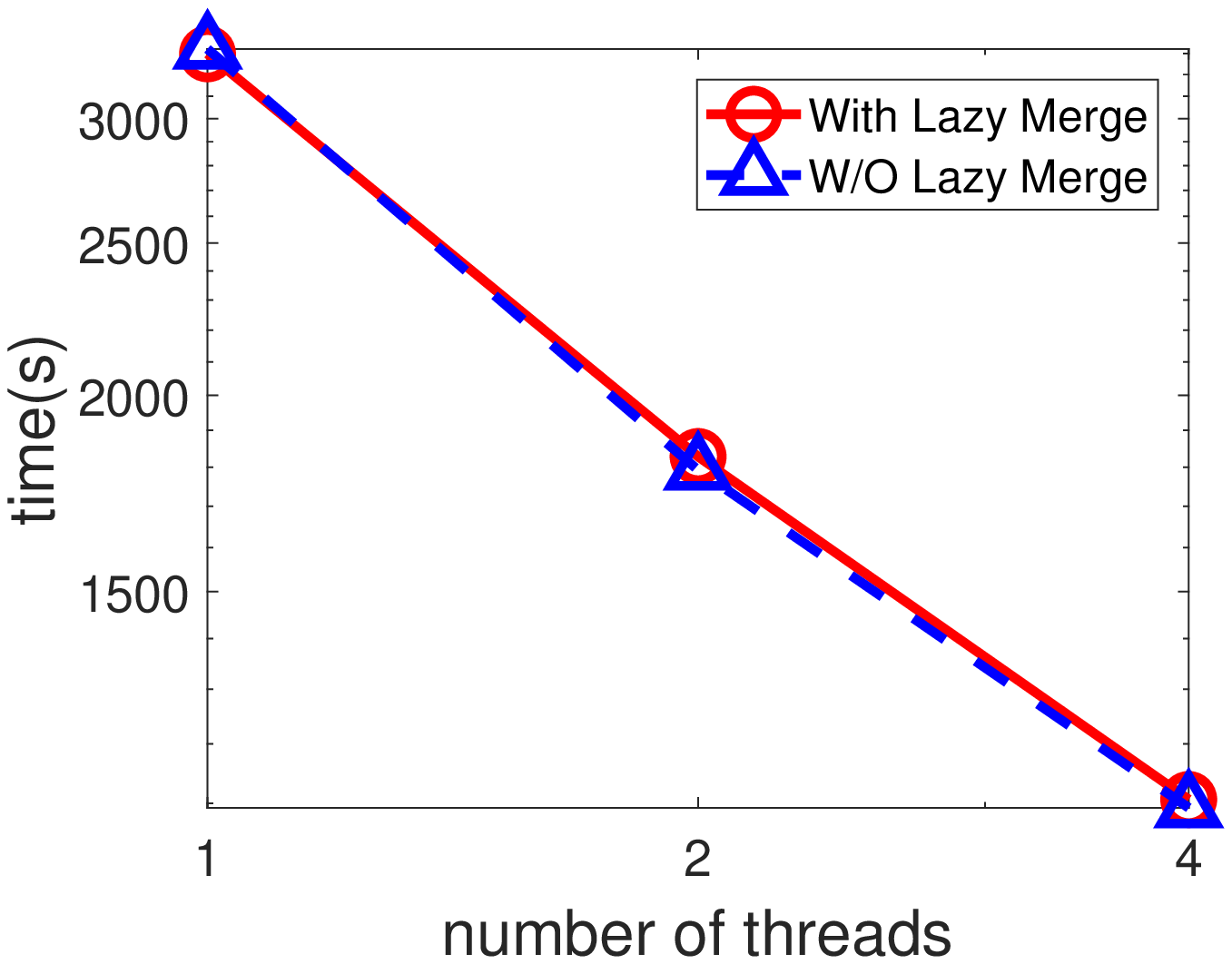}
\end{minipage}
}
\subfigure[Global parallelism.]{
\begin{minipage}[b]{0.22\textwidth}
\includegraphics[width=1.15\textwidth]{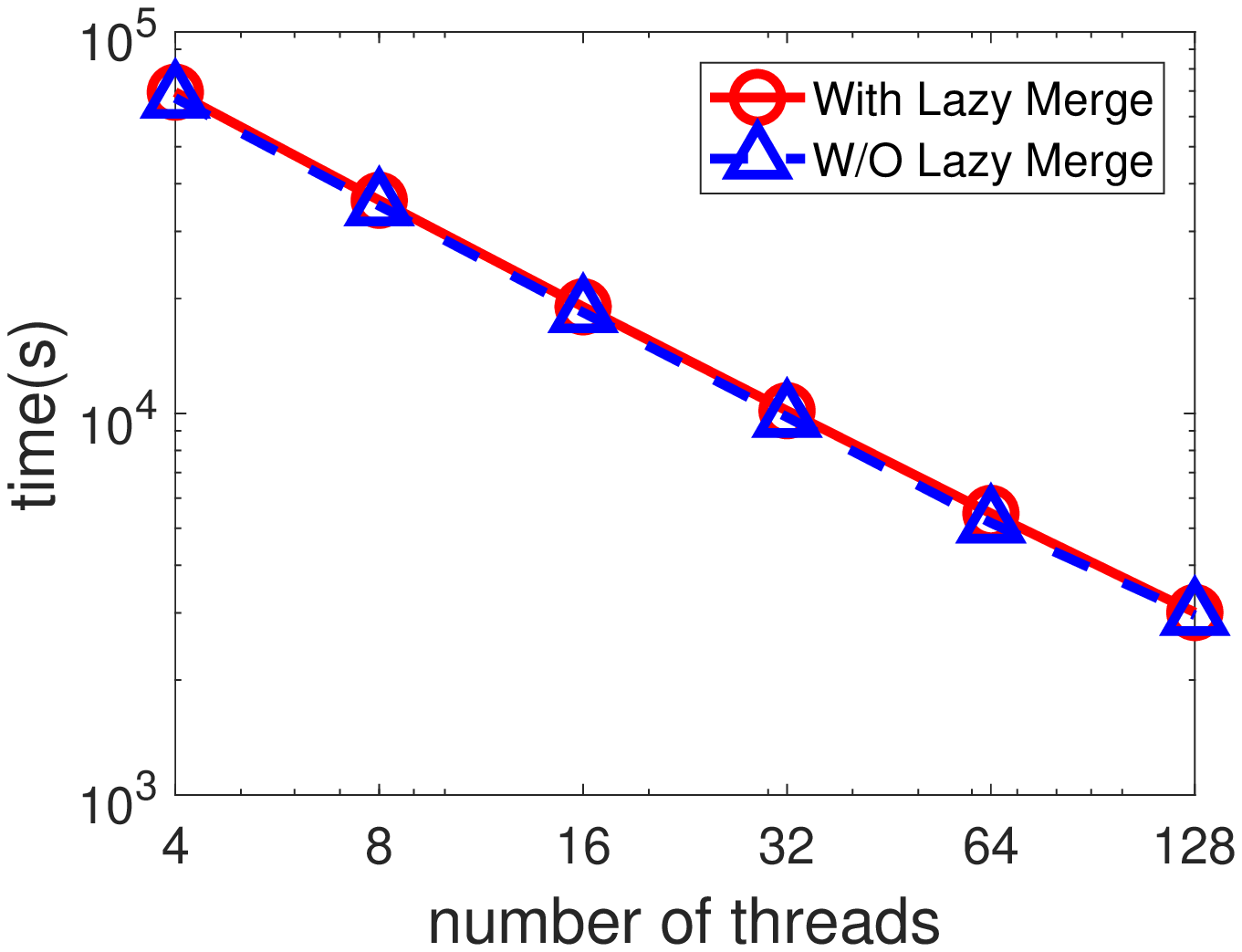}
\end{minipage}
}
\vspace{-1em}
\caption{Local parallelism and global parallelism.}
\label{fig:parallelism}
\end{figure}

As shown in Figure \ref{fig:parallelism}, if we use one machine with up to $4$ threads, the protocol almost shows a linear speedup. As the number of threads doubles, the computation time reduces to half. However, as we further increase the number of threads, we only see sub-linear speedup. We believe this is due to the small size of the dataset. In distributed computation experiments, we employed 4, 8, 16, 32, 64, and 128 threads, respectively. It is clear that at the beginning the protocol shows a linear speedup. While the number of threads reaches $64$, it switches to sub-linear speedup again due to the small size of dataset. In both local and global parallelism, we observe that the difference between with lazy merging and without lazy merging is too small to be observed. In other words, when we have enough computation power, lazy merging provides limited improvement, which is opposite to what we see in single-thread experiment.

\section{Conclusions}\label{sec:conclusion}

In this paper, we proposed a fully secure skyline protocol on encrypted data using two non-colluding cloud servers under the semi-honest model. It ensures semantic security in that the cloud servers knows nothing about the data including indirect data patterns, query, as well as the query result. In addition, the client and data owner do not need to participate in the computation. We also presented a secure dominance protocol which can be used by skyline queries as well as other queries. Furthermore, we demonstrated two optimizations, data partitioning and lazy merging, to further reduce the computation load. Finally, we presented our implementation of the protocol and demonstrated the feasibility and efficiency of the solution.  As for future work, we plan to optimize the communication time complexity to further improve the performance of the protocol.

\section*{Acknowledgement}

This research is supported in part by the Patient-Centered Outcomes Research Institute (PCORI) under award ME-1310-07058, the National Institute of Health (NIH) under award R01GM114612, and an NSERC Discovery grant.

\bibliographystyle{abbrv}
\bibliography{secureskyline}

\begin{thebibliography}{10}

\bibitem{DBLP:conf/fc/BaldimtsiO15}
F.~Baldimtsi and O.~Ohrimenko.
\newblock Sorting and searching behind the curtain.
\newblock In {\em {FC} 2015}, pages 127--146, 2015.

\bibitem{beimel2011secret}
A.~Beimel.
\newblock Secret-sharing schemes: a survey.
\newblock In {\em International Conference on Coding and Cryptology}, pages
  11--46. Springer, 2011.

\bibitem{DBLP:journals/cacm/Bentley80}
J.~L. Bentley.
\newblock Multidimensional divide-and-conquer.
\newblock {\em Commun. ACM}, 23(4):214--229, 1980.

\bibitem{DBLP:journals/jacm/BentleyKST78}
J.~L. Bentley, H.~T. Kung, M.~Schkolnick, and C.~D. Thompson.
\newblock On the average number of maxima in a set of vectors and applications.
\newblock {\em J. ACM}, 25(4):536--543, 1978.

\bibitem{DBLP:conf/icde/BorzsonyiKS01}
S.~B{\"o}rzs{\"o}nyi, D.~Kossmann, and K.~Stocker.
\newblock The skyline operator.
\newblock In {\em ICDE 2001}.

\bibitem{DBLP:conf/cikm/BotheCKV14}
S.~Bothe, A.~Cuzzocrea, P.~Karras, and A.~Vlachou.
\newblock Skyline query processing over encrypted data: An
  attribute-order-preserving-free approach.
\newblock In {\em PSBD@CIKM}, pages 37--43, 2014.

\bibitem{DBLP:journals/pvldb/BotheKV13}
S.~Bothe, P.~Karras, and A.~Vlachou.
\newblock eskyline: Processing skyline queries over encrypted data.
\newblock {\em {PVLDB}}, 6(12):1338--1341, 2013.

\bibitem{DBLP:conf/sigmod/ChanJTTZ06}
C.~Y. Chan, H.~V. Jagadish, K.-L. Tan, A.~K.~H. Tung, and Z.~Zhang.
\newblock Finding k-dominant skylines in high dimensional space.
\newblock In {\em SIGMOD Conference}, pages 503--514, 2006.

\bibitem{DBLP:conf/infocom/ChenLZZL16}
W.~Chen, M.~Liu, R.~Zhang, Y.~Zhang, and S.~Liu.
\newblock Secure outsourced skyline query processing via untrusted cloud
  service providers.
\newblock In {\em {INFOCOM} 2016}.

\bibitem{costanintel}
V.~Costan and S.~Devadas.
\newblock Intel sgx explained.
\newblock Technical report, Cryptology ePrint Archive, Report 2016/086, 20 16.
  http://eprint. iacr. org.

\bibitem{DBLP:conf/vldb/DellisS07}
E.~Dellis and B.~Seeger.
\newblock Efficient computation of reverse skyline queries.
\newblock In {\em VLDB}, pages 291--302, 2007.

\bibitem{DBLP:conf/icde/ElmehdwiSJ14}
Y.~Elmehdwi, B.~K. Samanthula, and W.~Jiang.
\newblock Secure k-nearest neighbor query over encrypted data in outsourced
  environments.
\newblock In {\em ICDE 2014}.

\bibitem{DBLP:conf/pet/ErkinFGKLT09}
Z.~Erkin, M.~Franz, J.~Guajardo, S.~Katzenbeisser, I.~Lagendijk, and T.~Toft.
\newblock Privacy-preserving face recognition.
\newblock In {\em {PETS}}, pages 235--253, 2009.

\bibitem{DBLP:journals/joc/FeigeFS88}
U.~Feige, A.~Fiat, and A.~Shamir.
\newblock Zero-knowledge proofs of identity.
\newblock {\em J. Cryptology}, 1(2):77--94, 1988.

\bibitem{DBLP:conf/stoc/Gentry09}
C.~Gentry.
\newblock Fully homomorphic encryption using ideal lattices.
\newblock In {\em STOC 2009}.

\bibitem{DBLP:books/cu/Goldreich2004}
O.~Goldreich.
\newblock {\em The Foundations of Cryptography - Volume 2, Basic Applications}.
\newblock Cambridge University Press, 2004.

\bibitem{DBLP:conf/stoc/GoldreichMW87}
O.~Goldreich, S.~Micali, and A.~Wigderson.
\newblock How to play any mental game or {A} completeness theorem for protocols
  with honest majority.
\newblock In {\em {ACM} Symposium on Theory of Computing}, pages 218--229,
  1987.

\bibitem{DBLP:conf/sigmod/HacigumusILM02}
H.~Hacig{\"{u}}m{\"{u}}s, B.~R. Iyer, C.~Li, and S.~Mehrotra.
\newblock Executing {SQL} over encrypted data in the database-service-provider
  model.
\newblock In {\em {SIGMOD} 2002}, pages 216--227, 2002.

\bibitem{DBLP:conf/eurocrypt/HaleviS15}
S.~Halevi and V.~Shoup.
\newblock Bootstrapping for helib.
\newblock In {\em {EUROCRYPT} 2015}, pages 641--670, 2015.

\bibitem{DBLP:conf/edbt/HashemKZ10}
T.~Hashem, L.~Kulik, and R.~Zhang.
\newblock Privacy preserving group nearest neighbor queries.
\newblock In {\em EDBT 2010}.

\bibitem{DBLP:conf/icde/HuXRC11}
H.~Hu, J.~Xu, C.~Ren, and B.~Choi.
\newblock Processing private queries over untrusted data cloud through privacy
  homomorphism.
\newblock In {\em ICDE 2011}.

\bibitem{DBLP:conf/uss/HuangEKM11}
Y.~Huang, D.~Evans, J.~Katz, and L.~Malka.
\newblock Faster secure two-party computation using garbled circuits.
\newblock In {\em {USENIX} 2011}, 2011.

\bibitem{DBLP:ucidata}
A.~Janosi, W.~Steinbrunn, M.~Pfisterer, and R.~Detrano.
\newblock Heart disease dataset,
  https://archive.ics.uci.edu/ml/datasets/heart+disease.
\newblock In {\em The UCI Archive 1998}.

\bibitem{DBLP:conf/compgeom/KirkpatrickS85}
D.~G. Kirkpatrick and R.~Seidel.
\newblock Output-size sensitive algorithms for finding maximal vectors.
\newblock In {\em Symposium on Computational Geometry}, pages 89--96, 1985.

\bibitem{DBLP:conf/vldb/KossmannRR02}
D.~Kossmann, F.~Ramsak, and S.~Rost.
\newblock Shooting stars in the sky: An online algorithm for skyline queries.
\newblock In {\em {VLDB} 2002}, 2002.

\bibitem{DBLP:journals/jacm/KungLP75}
H.~T. Kung, F.~Luccio, and F.~P. Preparata.
\newblock On finding the maxima of a set of vectors.
\newblock {\em JACM}, 1975.

\bibitem{DBLP:conf/cikm/Li0HRD12}
C.~Li, N.~Zhang, N.~Hassan, S.~Rajasekaran, and G.~Das.
\newblock On skyline groups.
\newblock In {\em CIKM}, pages 2119--2123, 2012.

\bibitem{DBLP:conf/icde/LiuZLLZZ15}
A.~Liu, K.~Zheng, L.~Li, G.~Liu, L.~Zhao, and X.~Zhou.
\newblock Efficient secure similarity computation on encrypted trajectory data.
\newblock In {\em ICDE}, pages 66--77, 2015.

\bibitem{DBLP:journals/pvldb/LiuXPLZ15}
J.~Liu, L.~Xiong, J.~Pei, J.~Luo, and H.~Zhang.
\newblock Finding pareto optimal groups: Group-based skyline.
\newblock {\em {PVLDB}}, 8(13):2086--2097, 2015.

\bibitem{DBLP:journals/ipl/LiuXX14}
J.~Liu, L.~Xiong, and X.~Xu.
\newblock Faster output-sensitive skyline computation algorithm.
\newblock {\em Inf. Process. Lett.}, 2014.

\bibitem{DBLP:conf/icde/LiuY0P17}
J.~Liu, J.~Yang, L.~Xiong, and J.~Pei.
\newblock Secure skyline queries on cloud platform.
\newblock In {\em {ICDE}}, pages 633--644, 2017.

\bibitem{DBLP:conf/icde/LiuY0PL18}
J.~Liu, J.~Yang, L.~Xiong, J.~Pei, and J.~Luo.
\newblock Skyline diagram: Finding the voronoi counterpart for skyline queries.
\newblock In {\em {ICDE}}, 2018.

\bibitem{DBLP:conf/cikm/LiuZXLL15}
J.~Liu, H.~Zhang, L.~Xiong, H.~Li, and J.~Luo.
\newblock Finding probabilistic k-skyline sets on uncertain data.
\newblock In {\em CIKM}, pages 1511--1520, 2015.

\bibitem{DBLP:conf/eurocrypt/Paillier99}
P.~Paillier.
\newblock Public-key cryptosystems based on composite degree residuosity
  classes.
\newblock In {\em Advances in Cryptology - {EUROCRYPT} '99}, pages 223--238,
  1999.

\bibitem{DBLP:journals/tods/PapadiasTFS05}
D.~Papadias, Y.~Tao, G.~Fu, and B.~Seeger.
\newblock Progressive skyline computation in database systems.
\newblock {\em {ACM} Trans. Database Syst.}, 30(1):41--82, 2005.

\bibitem{DBLP:journals/pvldb/PapadopoulosBP10}
S.~Papadopoulos, S.~Bakiras, and D.~Papadias.
\newblock Nearest neighbor search with strong location privacy.
\newblock {\em PVLDB}, 2010.

\bibitem{DBLP:conf/vldb/PeiJLY07}
J.~Pei, B.~Jiang, X.~Lin, and Y.~Yuan.
\newblock Probabilistic skylines on uncertain data.
\newblock In {\em VLDB}, pages 15--26, 2007.

\bibitem{DBLP:conf/icdcs/QiA08}
Y.~Qi and M.~J. Atallah.
\newblock Efficient privacy-preserving k-nearest neighbor search.
\newblock In {\em ICDCS 2008}.

\bibitem{DBLP:conf/ccs/SamanthulaHJ13}
B.~K. Samanthula, C.~Hu, and W.~Jiang.
\newblock An efficient and probabilistic secure bit-decomposition.
\newblock In {\em {ASIA} {CCS}}, pages 541--546, 2013.

\bibitem{DBLP:conf/sp/SongWP00}
D.~X. Song, D.~Wagner, and A.~Perrig.
\newblock Practical techniques for searches on encrypted data.
\newblock In {\em {IEEE} Symposium on Security and Privacy}, 2000.

\bibitem{DBLP:journals/jstsp/VeugenBHE15}
T.~Veugen, F.~Blom, S.~J.~A. de~Hoogh, and Z.~Erkin.
\newblock Secure comparison protocols in the semi-honest model.
\newblock {\em J. Sel. Topics Signal Processing}, 9(7):1217--1228, 2015.

\bibitem{DBLP:conf/sigmod/WongCKM09}
W.~K. Wong, D.~W. Cheung, B.~Kao, and N.~Mamoulis.
\newblock Secure knn computation on encrypted databases.
\newblock In {\em SIGMOD 2009}.

\bibitem{DBLP:conf/focs/Yao82b}
A.~C. Yao.
\newblock Protocols for secure computations (extended abstract).
\newblock In {\em {FOCS}}, pages 160--164, 1982.

\bibitem{DBLP:conf/icde/0002LX13}
B.~Yao, F.~Li, and X.~Xiao.
\newblock Secure nearest neighbor revisited.
\newblock In {\em ICDE 2013}.

\bibitem{DBLP:conf/icde/YiPBV14}
X.~Yi, R.~Paulet, E.~Bertino, and V.~Varadharajan.
\newblock Practical k nearest neighbor queries with location privacy.
\newblock In {\em ICDE 2014}.

\bibitem{DBLP:conf/cikm/YuQL0CZ17}
W.~Yu, Z.~Qin, J.~Liu, L.~Xiong, X.~Chen, and H.~Zhang.
\newblock Fast algorithms for pareto optimal group-based skyline.
\newblock In {\em {CIKM}}, pages 417--426, 2017.

\bibitem{DBLP:conf/edbt/ZhuMK14}
H.~Zhu, X.~Meng, and G.~Kollios.
\newblock Privacy preserving similarity evaluation of time series data.
\newblock In {\em EDBT}, pages 499--510, 2014.

\end{thebibliography}

\vspace{-1em}
\begin{IEEEbiography}[{\includegraphics[width=1in,height=1.25in,clip,keepaspectratio]{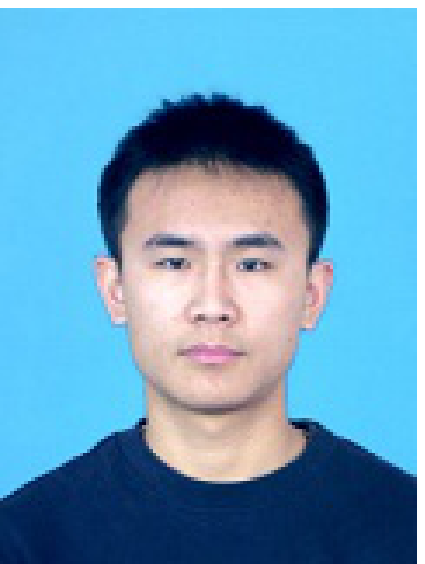}}]{Jinfei Liu}
is a joint postdoctoral research fellow at  Emory University and Georgia Institute of Technology. His research interests include skyline queries, data privacy and security, and machine learning. He has published over 20 papers in premier journals and conferences including VLDB, ICDE, CIKM, and IPL.
\end{IEEEbiography}

\vspace{-1em}
\begin{IEEEbiography}[{\includegraphics[width=1in,height=1.25in,clip,keepaspectratio]{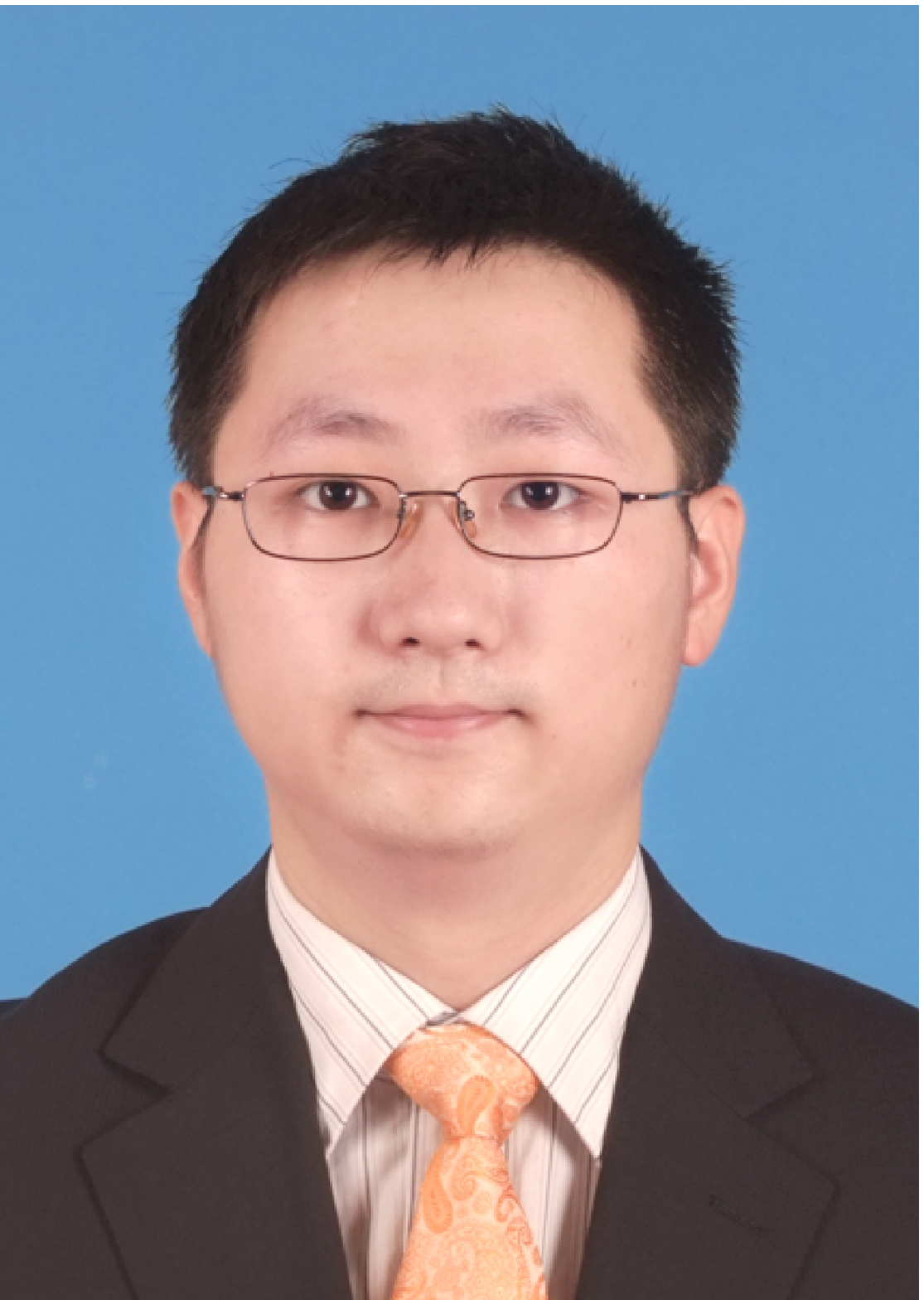}}]{Juncheng Yang}
is a master student in Emory University. His research interests include computer security, database, smart cache in storage and distributed system. He has published over 10 papers in premier conferences including ICDE and SoCC.
\end{IEEEbiography}

\vspace{-1em}
\begin{IEEEbiography}[{\includegraphics[width=1in,height=1.25in,clip,keepaspectratio]{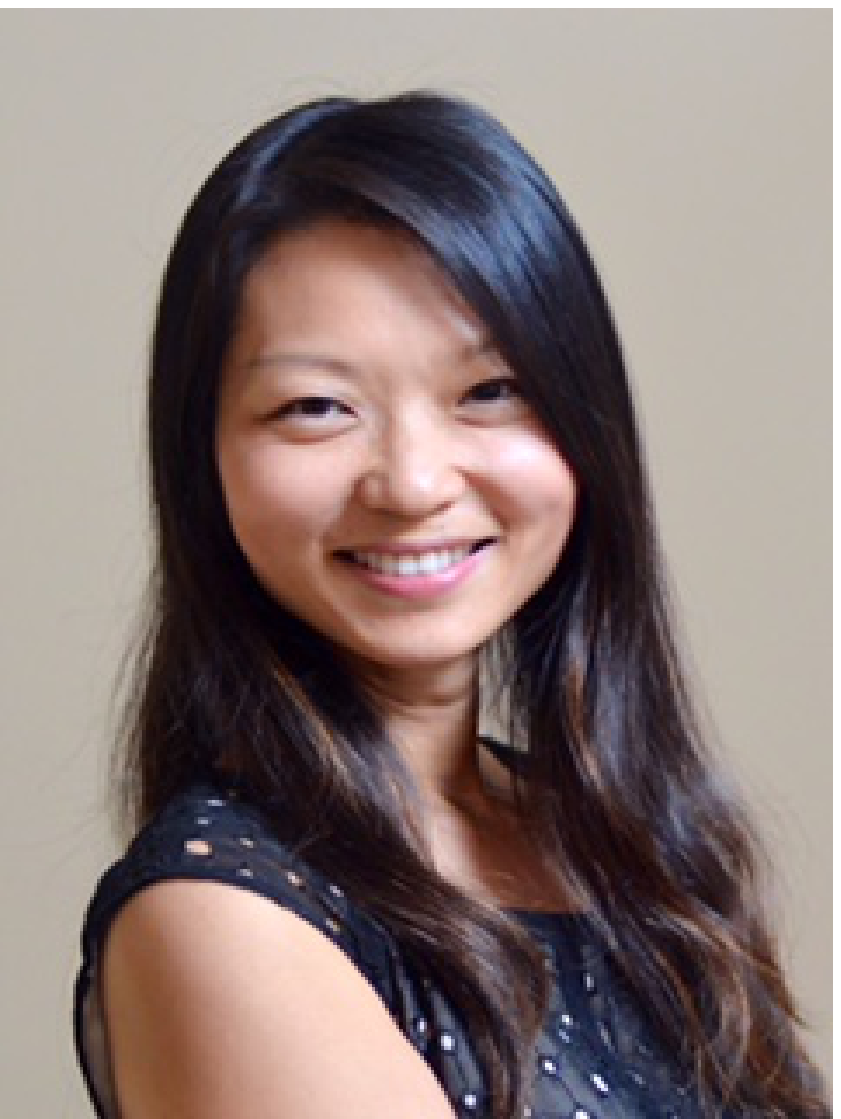}}]{Li Xiong}
is a Professor of Computer Science and Biomedical Informatics at Emory University. She conducts research that addresses both fundamental and applied questions at the interface of data privacy and security, spatiotemporal data management, and health informatics.  She has published over 100 papers in premier journals and conferences including TKDE, JAMIA, VLDB, ICDE, CCS, and WWW. She currently serves as associate editor for IEEE Transactions on Knowledge and Data Engineering (TKDE) and on numerous program committees for data management and data security conferences.
\end{IEEEbiography}

\vspace{-1em}
\begin{IEEEbiography}[{\includegraphics[width=1.5in,height=1.35in,clip,keepaspectratio]{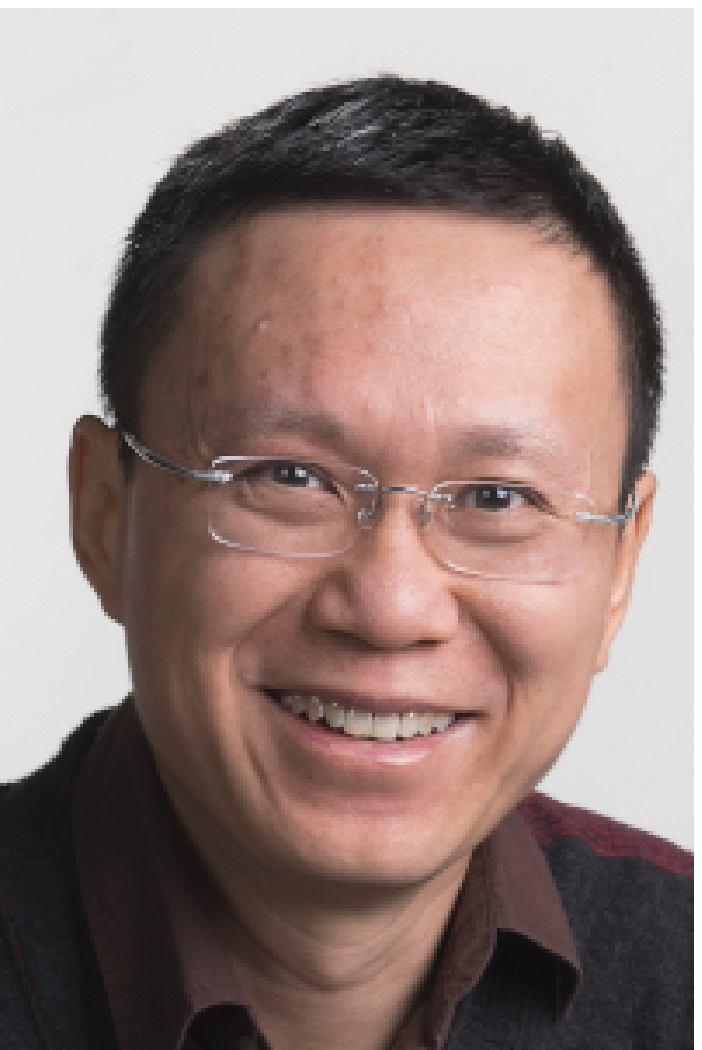}}]{Jian Pei}
is currently a Canada Research Chair (Tier 1) in Big Data Science, a Professor in the School of Computing Science at Simon Fraser University, Canada. He is one of the most cited authors in data mining, database systems, and information retrieval. Since 2000, he has published one textbook, two monographs and over 200 research papers in refereed journals and conferences, which have been cited by more than 77,000 in literature. He was the editor-in-chief of the IEEE Transactions of Knowledge and Data Engineering (TKDE) in 2013-2016, is currently a director of the Special Interest Group on Knowledge Discovery in Data (SIGKDD) of the Association for Computing Machinery (ACM). He is a Fellow of the ACM and of the IEEE.
\end{IEEEbiography}

\appendices

\section{Basic Security Subprotocols}
\partitle{Secure Multiplication (SM)}
Assume a cloud server $\mathcal{C}_1$ with encrypted input $E_{pk}(a)$ and $E_{pk}(b)$, and a cloud server $\mathcal{C}_2$ with the private key $sk$, where $a,b$ are two numbers not known to $\mathcal{C}_1$ and $\mathcal{C}_2$. The Secure Multiplication (SM) protocol \cite{DBLP:conf/icde/ElmehdwiSJ14} (based on the additively homomorphic property of Paillier) securely computes encrypted result of multiplication of $a,b$, $E_{pk}(a\times b)$, such that only $\mathcal{C}_1$ knows $E_{pk}(a\times b)$, and no information related to $a,b$ is revealed to $\mathcal{C}_1$ or $\mathcal{C}_2$.

\partitle{Secure Bit Decomposition (SBD)}
 Assume a cloud server $\mathcal{C}_1$ with encrypted input $E_{pk}(a)$ and a cloud server $\mathcal{C}_2$ with the private key $sk$, where $a$ is a number not known to $\mathcal{C}_1$ and $\mathcal{C}_2$. The Secure Bit Decomposition (SBD) protocol \cite{DBLP:conf/ccs/SamanthulaHJ13} securely computes encrypted individual bits of the binary representation of $a$, denoted as $\llbracket a \rrbracket=\langle E_{pk}((a)_B^{(1)}),...,E_{pk}((a)_B^{(l)})\rangle$, where $l$ is the number of bits, $(a)_B^{(1)}$ and $(a)_B^{(l)}$ denote the most and least significant bits of $a$, respectively.  At the end of the protocol, the output $\llbracket a \rrbracket$ is known only to $\mathcal{C}_1$ and no information related to $a$ is revealed to $\mathcal{C}_1$ or $\mathcal{C}_2$.

\subsection{Secure Boolean Operations}

\partitle{Secure OR (SOR)}
 Assume a cloud sever $\mathcal{C}_1$ with encrypted input $E_{pk}(\hat{a})$ and $E_{pk}(\hat{b})$, and a cloud server $\mathcal{C}_2$ with the private key $sk$, where $\hat{a}$ and $\hat{b}$ are two bits not known to $\mathcal{C}_1$ and $\mathcal{C}_2$. The Secure OR (SOR) protocol \cite{DBLP:conf/icde/ElmehdwiSJ14} securely computes encrypted result of the bit-wise OR of the two bits, $E_{pk}(\hat{a}\vee \hat{b})$, such that only $\mathcal{C}_1$ knows $E_{pk}(\hat{a}\vee \hat{b})$ and no information related to $\hat{a}$ and $\hat{b}$ is revealed to $\mathcal{C}_1$ or $\mathcal{C}_2$.

\partitle{Secure AND (SAND)}
 Assume a cloud server $\mathcal{C}_1$ with encrypted input $E_{pk}(\hat{a})$ and $E_{pk}(\hat{b})$, and a cloud server $\mathcal{C}_2$ with the private key $sk$, where $\hat{a}$ and $\hat{b}$ are two bits not known to $\mathcal{C}_1$ and $\mathcal{C}_2$. The goal of the SAND protocol is to securely compute encrypted result of the bit-wise AND of the two bits, $E_{pk}(\hat{a}\wedge \hat{b})$, such that only $\mathcal{C}_1$ knows $E_{pk}(\hat{a}\wedge \hat{b})$ and no information related to $\hat{a}$ and $\hat{b}$ is revealed to $\mathcal{C}_1$ or $\mathcal{C}_2$. We can simply use the secure multiplication (SM) protocol on the two bits.

\partitle{Secure NOT (SNOT)}
 Assume a cloud server $\mathcal{C}_1$ with encrypted input $E_{pk}(\hat{a})$ and a cloud server $\mathcal{C}_2$ with the private key $sk$, where $\hat{a}$ is a bit not known to $\mathcal{C}_1,\mathcal{C}_2$. The goal of the SNOT protocol is to securely compute the encrypted complement bit of $\hat{a}$, $E_{pk}(\neg \hat{a})$, such that only $\mathcal{C}_1$ knows $E_{pk}(\neg \hat{a})$ and no information related to $\hat{a}$ is revealed to $\mathcal{C}_1$ or $\mathcal{C}_2$. Secure NOT protocol can be easily implemented by $E_{pk}(1-\hat{a})=E_{pk}(1)E_{pk}(\hat{a})^{N-1}$.

\section{Disclosure of Binary based SMIN}\label{APP:1}
Given two numbers in binary representations, the idea of the Binary representation based SMIN protocol (BSMIN)\footnote{The SMIN protocol for $n$ values can be constructed by employing BSMIN for two values at a time in a hierarchical fashion as suggested in \cite{DBLP:conf/icde/ElmehdwiSJ14} or simply a linear fashion.} \cite{DBLP:conf/icde/ElmehdwiSJ14} is for $\mathcal{C}_1$ to randomly choose a boolean functionality $F$ (by flipping a coin), where $F$ is either $a>b$ or $b>a$, and then securely compute $F$ with $\mathcal{C}_2$, such that the output of $F$ is oblivious to both $\mathcal{C}_1$ and $\mathcal{C}_2$. Based on the output and chosen $F$, $\mathcal{C}_1$ computes $min(a,b)$ locally using homomorphic properties. More specifically, given the binary representation of the two numbers, for each bit, $\mathcal{C}_1$ computes an encrypted boolean output $W_i$ of the two bits based on $F$ (e.g., if F is $a>b$, $W_i = E_{pk}(1)$, if $(a)_B^{(i)} > (b)_B^{(i)}$ and $E_{pk}(0)$ otherwise) and an encrypted randomized difference between $(a)_B^{(i)}$ and $(b)_B^{(i)}$. This way, the order and difference of the two numbers are not disclosed to $\mathcal{C}_2$. However, when $a=b$, whatever $F$ is, we have $W_i=E_{pk}(0)$ for all bits. We can show that through the intermediate result (the encrypted randomized difference between $(a)_B^{(i)}$ and $(b)_B^{(i)}$, $\Gamma_i=E_{pk}(r_i)$ for $1\leq i\leq l$, the bit-wise XOR of $(a)_B^{(i)}$ and $(b)_B^{(i)}$, $G_i=E_{pk}(0)$ for $1\leq i\leq l$), $\mathcal{C}_2$ can determine $a$ equals to $b$.

\section{Disclosure of Perturbation based SMIN}\label{APP:2}
The Perturbation based SMIN protocol (PSMIN) \cite{DBLP:conf/edbt/ZhuMK14} assumes $\mathcal{C}_1$ has $E_{pk}(a)$ and $E_{pk}(b)$. $\mathcal{C}_1$ generates a set of $v$ random values uniformly from a certain range $\{r_1,...,r_v|r_1<r_i,i\geq 2\}$. $\mathcal{C}_1$ then sends a set of $2+v-1$ encrypted values $\{E_{pk}(a+r_1), E_{pk}(b+r_1), E_{pk}(x_2+r_2), ..., E_{pk}(x_v+r_v)\}$ to $\mathcal{C}_2$, where $x_i,i\geq 2$ are randomly chosen from $a,b$. The idea is that the smallest number, after being perturbed by $r_1$ (which is smaller than $r_i, i\geq 2$), will remain the smallest. The perturbation hides the order of the numbers to $\mathcal{C}_2$.  Although not mentioned by the original paper, we point out $\mathcal{C}_1$ also needs to shuffle the encrypted values before sending them to $\mathcal{C}_2$, otherwise the differences between the values will be disclosed to $\mathcal{C}_2$ after decryption. After decrypting those $2+v-1$ values, $\mathcal{C}_2$ takes the minimal $min$ and sends $E_{pk}(min)$ to $\mathcal{C}_1$. $\mathcal{C}_1$ computes $E_{pk}(min-r_1)$ as result. The security weakness of PSMIN is due to the fact that if two numbers are equal, their perturbed values remain equal. Since $\mathcal{C}_1$ sends $\{E_{pk}(a+r_1), E_{pk}(b+r_1), E_{pk}(x_2+r_2), ..., E_{pk}(x_v+r_v)\}$ to $\mathcal{C}_2$, $\mathcal{C}_2$ can learn two numbers are equal based on $a+r_1$ and $b+r_1$.

\section{Security Definition in the Semi-honest Model}
Considering the privacy properties above, we adopt the formal security definition from the multi-party computation setting under the semi-honest model \cite{DBLP:books/cu/Goldreich2004}. Intuitively, a protocol is secure if whatever can be computed by a party participating in the protocol can be computed based on its input and output only. This is formalized according to the simulation paradigm. Loosely speaking, we require that a party's view in a protocol execution to be simulative given only its input and output. This then implies that the parties learn nothing from the protocol execution. For the detailed and strict definition, please see \cite{DBLP:books/cu/Goldreich2004}.

\begin{theorem}{(\textbf{Composition Theorem})} \cite{DBLP:books/cu/Goldreich2004}. \label{def:comp}
If a protocol consists of subprotocols, the protocol is secure as long as the subprotocols are secure and all the intermediate results are random or pseudo-random.
\end{theorem}

In this work,  the proposed  secure skyline protocols are constructed based on  a sequential composition of subprotocols. 
To formally prove the security under the semi-honest model, according to the composition theorem given in Theorem \ref{def:comp}, one needs to show that the simulated view of each subprotocol was computationally indistinguishable from the actual execution view and the protocol produces random or pseudo-random shares as intermediate results.

\section{Paillier Cryptosystem}
We use the Paillier cryptosystem \cite{DBLP:conf/eurocrypt/Paillier99} as the encryption scheme in this paper and briefly describe Paillier's additive homomorphic properties which will be used in our protocols.

$\bullet$ Homomorphic addition of plaintexts:

\begin{displaymath}
D_{sk}(E_{pk}(a) \times E_{pk}(b)~mod~N^2)=(a+b)~mod~N
\end{displaymath}
$~~~~~\bullet$ Homomorphic multiplication of plaintexts:
\begin{displaymath}
D_{sk}(E_{pk}(a)^b~mod~N^2)=a\times b~mod~N
\end{displaymath}
It is easy to see that the Paillier cryptosystem is additively homomorphic and we can compute a new probabilistic encrypted $E_{pk}(a)$ given an encrypted $E_{pk}(a)$ without knowing the private key $sk$.


\end{document}